\documentclass[reprint,twocolumn,aps,prb,superscriptaddress]{revtex4-2}
\usepackage[T1]{fontenc}
\usepackage[utf8]{inputenc}
\usepackage{lmodern}
\usepackage{graphicx}
\usepackage{amsmath}
\usepackage{xcolor}
\usepackage{float}
\usepackage{siunitx}
\usepackage{hyperref}
\usepackage{braket}
\usepackage{amssymb,amsmath,amsthm, array}
\usepackage{mdframed}
\usepackage{systeme,mathtools}
\usepackage{yfonts}
\usepackage{comment}
\usepackage{bm}

\DeclareUnicodeCharacter{0301}{-}

\usepackage{tikz}
\usepackage{pgfplots}
\pgfplotsset{compat=1.18}
\definecolor{tealc}{HTML}{0F6E56}
\definecolor{coralc}{HTML}{B0451D}
\definecolor{grayc}{HTML}{5F5E5A}

\begin{document}
\renewcommand{\vec}[1]{\mathbf{#1}}
\newcommand{\ii}{\mathrm{i}}
\def\ya#1{{\color{orange}{#1}}}

\title{Hopf bifurcation and stochastic spiking in an antiferromagnetic FitzHugh--Nagumo normal form}

\author{D. Maroulakos}
\affiliation{Doctoral School at the  University of Rzesz\'ow, 35-317 Rzesz\'ow, Poland}

\author{A. Wal}
\affiliation{Institute of Physics, Faculty of Exact and Technical Sciences, University of Rzesz\'ow, 35-317 Rzesz\'ow, Poland}

\author{I. Tralle}
\affiliation{Institute of Physics, Faculty of Exact and Technical Sciences, University of Rzesz\'ow, 35-317 Rzesz\'ow, Poland}

\author{S. K. Mishra}
\affiliation{Department of Physics, Indian Institute of Technology (Banaras Hindu University) Varanasi - 221005, India}

\author{L. Chotorlishvili}
\affiliation{Department of Physics and Medical Engineering, Rzesz\'ow University of Technology, 35-959 Rzesz\'ow, Poland}

\date{\today}

\begin{abstract}

Antiferromagnets offer ultrafast, stray-field-free dynamics that are attractive for neuromorphic spintronic devices. Here we analyze an antiferromagnetic spin-Hall nano-oscillator in the overdamped regime and derive a reduced set of equations for the N\'eel-vector dynamics constrained to the unit sphere. For spin polarization along the easy axis, the model reduces to an asymmetric rotator, for which analytic solutions and the associated spin-pumping signal are obtained in selected limits. We further show that near a suitable operating point, the projected dynamics can be transformed into a local FitzHugh-Nagumo normal form. The resulting mapping identifies the effective fast variable, recovery variable, bias current, and Hopf condition in terms of magnetic material parameters. We finally extend the reduced model to an It\^o stochastic FitzHugh-Nagumo equation driven by spin-pumping input and additive thermal or electronic fluctuations. The stochastic phase portrait shows that the deterministic nullcline geometry organizes noisy spike cycles and produces controlled spike-time variability. These results provide a minimal analytic framework for AFM-based spiking-neuron elements and suggest design criteria for future neuromorphic spintronic devices..

\end{abstract}

\maketitle

\section{Introduction}

Due to the complexity of the human brain, constructing detailed, realistic models requires certain simplifications. However, the simplified model should mimic and adopt the brain's core design principles, i.e., billions of neurons and trillions of synapses should be described by relatively simple, computationally accessible models  \cite{zhu2020comprehensive,wang2017nanoionics,gupta2025toward,aimone2022review}. The main goal of neuromorphic computing is to emulate the operational principles of biological neural networks and overcome the limitations of conventional von Neumann architectures, particularly with respect to energy efficiency, parallelism, and scalability \cite{mead2002neuromorphic}. In the von Neumann architecture, the central processing unit (CPU), a separate memory unit (RAM), and a connecting bus are implemented. This leads to the ``von Neumann bottleneck''. In essence, the CPU wastes time and energy by repeatedly fetching data from memory, even for simple tasks. Neuromorphic computing is thought capable to solve this problem. Hardware realizations of neuromorphic systems rely on physical devices capable of exhibiting nonlinear dynamics, memory, and adaptability, properties that have driven intense research into emerging materials and spintronic platforms \cite{grollier2016spintronic,torrejon2017neuromorphic}. Among these, antiferromagnetic (AFM) materials have recently emerged as promising, largely unexplored candidates for neuromorphic computing.
Magnetic materials are very promising in the neuromorphic context and are thought considered  to serve as prototype platforms in the futuristic in future  neuromorphic devices \cite{godinho2024antiferromagnetic,han2024neuromorphic,zhou2021prospect,fukami2018perspective,sulymenko2018ultra,ojha2024neuromorphic,
fukami2020antiferromagnetic,kurenkov2019artificial,d6z2-t1sm,PhysRevApplied.19.064010,PhysRevB.108.184411,PhysRevApplied.21.040503}. However, it is worth mentioning that antiferromagnets (AFMs) possess intrinsic advantages over ferromagnets. These advantages are widely discussed in the spintronic literature, and they hold in the neuromorphic context as well. To name but a few: the absence of stray fields, robustness against external magnetic perturbations, and ultrafast spin dynamics in the terahertz regime \cite{jungwirth2016antiferromagnetic}. These characteristics make AFMs attractive for high-speed operation platforms, desirable for neuromorphic architectures. Moreover, the staggered spin order in AFMs enables rich internal dynamics that can be harnessed for information processing beyond binary logic \cite{gomonay2014spintronics}. Given the above arguments, AFMs are attractive for implementation on diverse high-speed operation platforms for neuromorphic architectures. Several theoretical and experimental studies have demonstrated that the Néel vector in AFMs can be switched by spin-polarized current. Besides, spin–orbit torque effects play an important role in AFMs and lead to effects that mimic artificial synapses and neurons \cite{wadley2016electrical,bodnar2018writing}. In particular, electrically controlled AFM devices have been proposed as memristive elements capable of synaptic weight storage and plasticity \cite{PhysRevLett.117.087203}. Such functionality is central to neuromorphic computation, where learning and adaptation are encoded in analog changes of device conductance. Beyond static memory effects, the intrinsic dynamics of AFMs offer additional opportunities for neuromorphic computing paradigms such as reservoir computing and spiking neural networks \cite{PhysRevApplied.19.024063,han2024neuromorphic,bradley2024pattern,zhou2021prospect}.
\begin{figure}[htbp]
    \centering
    \includegraphics[width=\columnwidth]{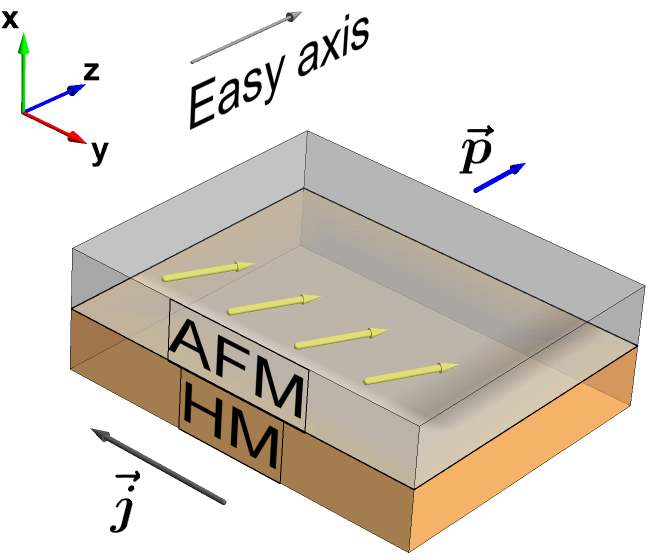}
    \caption{Schematic representation of the setup: an antiferromagnetic material placed on a heavy metal. The electric current in heavy metal flows in $-\bm {y} $ direction. The unit vector $\bm {p} $ defines polarization of the spin transfer torque. When $\bm {p}$ is polarized along $\bm z$ axis,, in the overdamped regime, magnetization dynamics reduces to the “FitzHugh–Nagumo model. }
    \label{fig:1}
\end{figure}
The magnetization dynamics in AFMs are inherently nonlinear. The nonlinear response of AFM order parameters to external driving fields or spin-polarized currents makes it an illustrative example of the application of fundamental mathematical methods of dynamical systems theory. Besides, dynamics in AFMs are much faster than in the FMs. This is already a practical advantage for the faster processing of temporal information. These properties suggest that AFM-based neuromorphic devices could operate in regimes inaccessible to existing hardware. From a materials perspective, a wide range of metallic and insulating antiferromagnets—including Mn-based alloys, oxides, and van der Waals materials—are being actively investigated for spintronic and neuromorphic applications \cite{roldan2022spiking,cao20212d}. Despite this progress, AFM neuromorphic computing remains in its early stages, and several challenges persist. In this work, we show that, under an overdamped approximation and near a suitable operating point, the projected AFM Néel-vector dynamics admits a local normal-form reduction to FitzHugh-Nagumo-type equations. This provides an analytic route toward AFM-based spiking neuromorphic elements. Our results establish a minimal, analytically tractable framework for AFM-based neuromorphic devices.

\section{Model}
We consider AFM material placed on the Heavy Metal layer as shown in Fig.\ref{fig:1}. The charge current $\textbf{j}$ is flowing in the heavy metal (HM, e.g., Pt). Electric current passing through the Heavy Metal is converted into the spin current via the Spin Hall Effect. Spin current exerts spin-transfer torque (STT) on the magnetization dynamics in the AFM system. The easy axis of magnetization is oriented along the $\textbf{z}$ axis. The spin polarization direction of the injected spin current is given by $\textbf{p}=\textbf{j}\times\textbf{n}$, where $\textbf{n}$ is the normal to the HM/AFM interface. When $\textbf{p}=\textbf{z}$, then spin current tries to rotate the Néel vector $\bm{I}=\textbf{m}_1-\textbf{m}_2$ around the easy axis, leading to the AFM nano-oscillator regime. Here, $\textbf{m}_1$ and $\textbf{m}_2$ are magnetization vectors of two AFM sublattices. For more details, we refer to \cite{PhysRevApplied.18.024047}. To proceed further, we introduce the action of the system:
\begin{eqnarray}\label{the action}
S=\int\mathcal L \left(\bm I, \dot{\bm I}\right)dt.
\end{eqnarray}
The action introduced here leads to the \textit{Euler-Lagrange-Rayleigh} equation projected on the sphere and tangential dynamics
\begin{eqnarray}\label{equations of Lagrange}
&& \bm I\times\frac{d}{dt}\frac{\partial\mathcal{L}}{\partial\dot{\textbf{I}}}=\bm I\times\frac{\partial\mathcal{L}}{\partial\textbf{I}}-\bm I\times\frac{\partial\mathcal{R}}{\partial\dot{\textbf{I}}},
\end{eqnarray}
where $\bm{I}=(I_x, I_y, I_z)$.  
Transition to the dimensionless equations can be done through the transformations: $\tau_m\equiv\omega_{\rm ex}\tau_m$, $\omega_0\equiv \omega_0/\omega_{\rm ex}$, $\omega_{\rm ip}\equiv \omega_{\rm ip}/\omega_{\rm ex}$, $\tau\equiv\tau/\omega_{\rm ex}$. Then the dimensionless Lagrange and Rayleigh functions read
\begin{eqnarray}\label{2appendix equations of Lagrange}
&& \mathcal L=\frac{1}{2}\dot{\bm I}^{\,2}-\frac{1}{2}
\left[\omega_0^2 (I_x^2 + I_y^2)+\omega_{\rm ip}^2 I_y^2
\right],\nonumber\\
&&\mathcal R=\frac{\alpha }{2}\dot{\bm I}^{\,2}+\tau\,\bm p \cdot (\bm I \times \dot{\bm I}),\nonumber\\
&&w_a=\frac{1}{2}\left[\omega_0^2 (I_x^2 + I_y^2)+\omega_{\rm ip}^2 I_y^2\right].
\end{eqnarray}
Here $\alpha$ is the Gilbert damping, $M_s$ is the sublattice magnetization, $A$ is the exchange stiffness, $H_{\mathrm{ex}}$ is the exchange field, $K$ is the easy-axis anisotropy constant, while  
$K_{\mathrm{ip}}$ corresponds to the in-plane (secondary) anisotropy. The exchange frequency is described by $\omega_{ex}$. The easy-axis magnon gap: $\omega_0=\gamma\sqrt{\frac{H_{\rm ex}K}{M_s}}$, $\omega_{\rm ip}=\gamma\sqrt{\frac{H_{\rm ex}K_{\rm ip}}{M_s}}$ and the torque coefficient $\tau=\sigma j$.  Our purpose is to derive the set of equations that conserve $\bm I^2=1$ in the overdamped regime. Being interested in tangential dynamics, we uncross the left-hand side of Eq.(\ref{equations of Lagrange}), neglect the normal term $|\dot{\bm I}|^2\bm I$ and second derivatives. Then, after a little algebra, we obtain
\begin{eqnarray}\label{explicit in components}
&&\alpha\dot I_x=-\left(\omega_0^2+\omega_{ip}^2\right) I_y I_z+\tau\left[ p_x-(p_x I_x+p_z I_z)I_x\right] ,\nonumber\\
&&\alpha\dot I_y=\,\omega_0^2\, I_x I_z-\tau\,(p_x I_x+p_z I_z)\,I_y,\\
&&\alpha\dot I_z=\omega_{ip}^2\, I_x I_y+\tau\left[ p_z-(p_x I_x+p_z I_z)I_z\right].\nonumber
\end{eqnarray}
We note that Eq.(\ref{explicit in components}) is the Euler–Lagrange equation projected onto the tangent space, and it describes dynamics on the unit sphere and preserves the normalization condition $\bm I^2=\rm{const}$. Besides, nonlinearity of the equations leads to the amplitude dependent oscillation frequency and nonlinear resonance
\cite{chotorlishvili2010quantum,PhysRevE.68.026216,PhysRevE.70.026219,PhysRevE.71.056211}. We proceed with the expression for the current: 
\begin{eqnarray}\label{solution4}
\mathbf J_s(t)= g\,\bm I(t)\times \dot{\bm I}(t).
\end{eqnarray} 
Since $|\bm I|=1$, one has $\bm I\cdot\dot{\bm I}=0$, and therefore
\begin{eqnarray}\label{solution5}
|\mathbf J_s(t)|= g\,|\dot{\bm I}(t)|.
\end{eqnarray} 
The spin-pumping signal is converted into an effective synaptic current
\begin{eqnarray}\label{solution6}
I_{\mathrm{syn}}(t)= \eta\,|\mathbf J_s(t)|= \eta g\,|\dot{\bm I}(t)| ,
\end{eqnarray} 
where $\eta$ is an effective spin-to-charge conversion coefficient. In what follows, we explore several cases for Eq.(\ref{explicit in components})  and obtain analytic solutions. The equation for the membrane potential reads:
\begin{eqnarray}\label{solution6and membrane}
&&\tau_m\frac{dv}{dt}=-(v-v_{\rm rest})+\beta \left\vert\frac{d\bm I}{dt}\right\vert,    
\end{eqnarray}
where $\beta=R_m\eta g\omega_{\rm ex}/V_0\approx 3$, $\tau_m^{\rm phys}=R_mC_m$, $R_m$ is the membrane (leak) resistance, $C_m$ is the membrane capacitance,  $\tau_m^{\rm phys}=1$ns, $\tau_m=30$. Before discussing the analytic results we define the values of parameters considered hereafter: The amplitude of the electric current  $j=\frac{\gamma H_{\rm ex}}{5\sigma}$, the gyro-magnetic ratio $\gamma=1,76\times 10^{11}$reads$^{-1}$T$^{-1}$. For synthetic AFMs, e.g., CoFeB/Ru/CoFeB, two identical layers, with $t_{AFM}=10$nm thickness, $M_s=1.2\cdot10^6$A/m, saturation magnetization, weak coupling interlayer exchange energy per area $J=1$mJ/m$^2$, $H_{\rm ex}=\frac{2J}{M_st_{AFM}}=0.17\text{T}$ leads to the exchange frequency $\omega_{\rm ex}=\gamma H_{\rm ex}=3\cdot10^{10}s^{-1}$, while the anisotropy constants $K=K_{\rm ip}=0.01\mu_0M_sH_{\rm ex}$. The rest of the parameters $\omega_0=\gamma\sqrt{H_{\rm ex}K/M_s}=0.1\omega_{\rm ex}$,  $\omega_{\rm ip}=\gamma\sqrt{H_{\rm ex}K_{\rm ip}/M_s}=0.1\omega_{\rm ex}$,  $j=6\cdot10^{13}$A/m$^2$, $\tau_{\rm phys}=\sigma j=6\cdot 10^{7}$s$^{-1}$, $\tau=\tau_{\rm phys}/\omega_{\rm ex}=0.002$, $\sigma =10^{-4}$m$^2$/(sA), and dimensionless notations $t\equiv t_{\rm phys}\, \omega_{\rm ex}$, $\alpha=0.1$, $\omega_0\equiv\omega_0/\omega_{\mathrm{ex}}=0.1$, $\omega_{\rm ip}\equiv\omega_{\rm ip}/\omega_{\mathrm{ex}}=0.1$.

We note that the pin-pumping signal converted into an effective synaptic current Eq.(\ref{solution6}) influences dynamics and spikes of the membrane potential  Eq.(\ref{solution6and membrane}). On the other hand, in what follows we demonstrate that Néel dynamics itself, under the certain conditions reduces locally to an FitzHugh–Nagumo-like spiking model.

\section{Solution of the overdamped problem}

In the limit of the overdamped motion $1/\omega_{\rm ex}\ll\alpha$, Eq.(\ref{explicit in components}) reduces to the problem of an asymmetric rotator. With \(p_x=0\), the overdamped equations of motion obtained from Eq.~(\ref{explicit in components}) take the form:
\begin{eqnarray}\label{solution1}
&&\alpha \dot I_x =- A\, I_y I_z - \tau p_z I_z I_x,\nonumber\\
&&\alpha \dot I_y =\phantom{-} B\, I_x I_z - \tau p_z I_z I_y,\\
&&\alpha \dot I_z =\phantom{-} C\, I_x I_y + \tau p_z (1-I_z^2),\nonumber
\end{eqnarray}
where the coefficients are defined as $A=\left(\omega_0^2+\omega_{ip}^2\right)$, $B=\omega_0^2$, $C=\omega_{\rm ip}^2$, so that \(A=B+C\).
We introduce the cylindrical variables on the unit sphere: $I_x=\rho\cos\psi$, $I_y=\rho\sin\psi$, $I_z=z$, $\rho^2=1-z^2$. Then, the equations of motion reduce to:
\begin{eqnarray}\label{solution2}
&&\alpha \dot\psi = z\bigl(B+C\sin^2\psi\bigr),\\
&&\alpha \dot z = (1-z^2)\left(\frac{C}{2}\sin 2\psi+\tau p_z\right).\nonumber
\end{eqnarray}
To solve Eq.(\ref{solution1}), we note that $z$ is slow variable, while $\psi(t)$ is fast variable $|\dot{z}/z|\ll|\dot{\psi}|$. Considering $I_z(0)$ as adiabatic integral, when integrate equation for $\dot{\psi}(t)$ we replace $z=I_z(0)$. Obtained solution $\psi(t)$ we insert in the second equation and integrate $\dot{z}(t)$. Then from Eq.(\ref{solution2}) we obtain an approximate solution:
\begin{eqnarray}\label{solution3}
&&\psi(t)=\arctan\bigg\lbrace
\sqrt{\frac{B}{B+C}}\,
\tan\!\bigg[
\arctan\!\Big(\sqrt{\tfrac{B+C}{B}}\tan\psi_0\Big)\nonumber\\ 
&&+\sqrt{B(B+C)}\,\frac{z_0}{\alpha}\,t
\bigg]\bigg\rbrace,
\end{eqnarray}
\begin{eqnarray}\label{+solution3}
&&z(t)=\tanh\!\bigg[
\operatorname{tanh}^{-1} (z_0)
+\frac{\tau p_z}{\alpha}\,t\nonumber\\
&&+\frac{1}{2z_0}
\ln\!\left(
\frac{B+C\sin^2\psi(t)}
     {B+C\sin^2\psi_0}
\right)
\bigg].
\end{eqnarray}
Then, using the parametrization
\begin{eqnarray}\label{solution9}
I_x=\rho\cos\psi,\;
I_y=\rho\sin\psi,\;
I_z=z,
\end{eqnarray} 
with $\rho^2=1-z^2$, the squared velocity of the order parameter is
\begin{eqnarray}\label{solution10}
|\dot{\mathbf I}|^2
= \dot I_x^2+\dot I_y^2+\dot I_z^2
= \frac{\dot z^{\,2}}{1-z^2}
+ (1-z^2)\dot\psi^{\,2}.
\end{eqnarray}
Hence the spin-pumping current can be written as
\begin{eqnarray}\label{solution11}
|\mathbf J_s(t)|
= g\sqrt{
\frac{\dot z^{\,2}}{1-z^2}
+ (1-z^2)\dot\psi^{\,2}
} \, .
\end{eqnarray} 
In the overdamped regime with $p_x=0$, Eq.~(\ref{solution2}) gives
\begin{eqnarray}\label{solution12}
&&\dot\psi
= \frac{z}{\alpha}\bigl(B+C\sin^2\psi\bigr),\nonumber\\
&&\dot z= \frac{1-z^2}{\alpha}
\left(\frac{C}{2}\sin 2\psi+\tau p_z\right).
\end{eqnarray}
Substituting Eq.~(\ref{solution12}) into Eq.~(\ref{solution11}), we obtain:
\begin{eqnarray}\label{solution13}
|\mathbf J_s(t)|
= \frac{g}{\alpha}
\sqrt{G_0(z,\psi)+G_1(z,\psi)} \, .
\end{eqnarray} 
where $G_0(z,\psi)=(1-z^2)z^2\bigl(B+C\sin^2\psi\bigr)^2$, and $G_1(z,\psi)=
(1-z^2)\left(\frac{C}{2}\sin 2\psi+\tau p_z\right)^2$. The corresponding synaptic current driving the neuron is therefore
\begin{eqnarray}\label{solution14}
I_{\mathrm{syn}}(t)
= \frac{\eta g}{\alpha}
\sqrt{G_0(z,\psi)+G_1(z,\psi)}  \, .
\end{eqnarray}
In the case $C=0$ (i.e.\ $\omega_{ip}=0$) the set of equations Eq.(\ref{solution1}) simplifies:
\begin{eqnarray}\label{1simplifies}
&&\alpha \dot I_x = -B I_y I_z - \tau p_z I_z I_x, \nonumber\\
&&\alpha \dot I_y = \phantom{-}B I_x I_z - \tau p_z I_z I_y, \\
&&\alpha \dot I_z = \tau p_z (1-I_z^2) \, .\nonumber
\end{eqnarray}
The equation for $I_z$ is decoupled and reads
\begin{eqnarray}\label{2simplifies}
\dot I_z = \kappa(1-I_z^2),
\qquad
\kappa=\frac{\tau p_z}{\alpha}.
\end{eqnarray}
The solution reads:
\begin{eqnarray}\label{3simplifies}
I_z(t)=\tanh\!\left(\kappa t+\operatorname{arctanh} z_0\right),
\qquad
z_0=I_z(0).
\end{eqnarray}
We introduce the cylindrical coordinates on the unit sphere 
\begin{eqnarray}\label{4simplifies}
&&I_x=\rho\cos\psi,\qquad
I_y=\rho\sin\psi,\qquad
I_z=z,\\
&&\rho^2=1-z^2,\nonumber
\end{eqnarray}
Eq.~(\ref{solution1}) with $C=0$ gives
\begin{eqnarray}\label{5simplifies}
\alpha \dot\psi = B z.
\end{eqnarray}
Substituting the solution for $z(t)$ and integrating yields
\begin{eqnarray}\label{6simplifies}
&&\psi(t)=\psi_0+\frac{B}{\alpha}\int_0^t z(s)\,ds =\nonumber\\
&&\psi_0+\frac{B}{\alpha\kappa}\ln\!\left[
\frac{\cosh\!\left(\kappa t+\operatorname{arctanh} z_0\right)}
     {\cosh\!\left(\operatorname{arctanh} z_0\right)}
\right],
\end{eqnarray}
or equivalently
\begin{eqnarray}\label{7simplifies}
\psi(t)
=
\psi_0+\frac{B}{\tau p_z}
\ln\!\left[
\frac{\cosh\!\left(\kappa t+\operatorname{arctanh} z_0\right)}
     {\cosh\!\left(\operatorname{arctanh} z_0\right)}
\right].
\end{eqnarray}
\paragraph*{Explicit solution for $I_x(t)$ and $I_y(t)$.}
Since $\rho(t)=\sqrt{1-z^2(t)}=\operatorname{sech}\!\left(\kappa t+\operatorname{artanh} z_0\right)$,
we obtain
\begin{align}\label{8simplifies}
&I_x(t) = \operatorname{sech}\!\left(\kappa t+\operatorname{arctanh} z_0\right)\cos\psi(t),\nonumber\\
&I_y(t) = \operatorname{sech}\!\left(\kappa t+\operatorname{arctanh} z_0\right)\sin\psi(t),\\
&I_z(t) = \tanh\!\left(\kappa t+\operatorname{arctanh} z_0\right).\nonumber
\end{align}
This solution satisfies  the identity $I_x^2+I_y^2+I_z^2=1$.
\paragraph*{Spin-pumping current}
From Eq.~(\ref{solution5}) the synaptic current is
\begin{eqnarray}\label{9simplifies}
I_{\mathrm{syn}}(t)=\eta g\,|\dot{\mathbf I}(t)|.
\end{eqnarray}
Using
\begin{eqnarray}\label{10simplifies}
|\dot{\mathbf I}|^2
=\frac{\dot z^2}{1-z^2}+(1-z^2)\dot\psi^{\,2},
\end{eqnarray}
together with
\begin{eqnarray}\label{11simplifies}
\dot z=\kappa(1-z^2),\qquad
\dot\psi=\frac{B}{\alpha}z,
\end{eqnarray}
we obtain
\begin{align}\label{12simplifies}
|\dot{\mathbf I}(t)|&=
\operatorname{sech}\!\left(\kappa t+\operatorname{arctanh} z_0\right)\times\nonumber\\
&\times \sqrt{\kappa^2+\left(\frac{B}{\alpha}\right)^2
\tanh^2\!\left(\kappa t+\operatorname{arctanh} z_0\right)
} \, .
\end{align}
Therefore,
\begin{align}\label{13simplifies}
I_{\mathrm{syn}}(t)&=\eta g\,\operatorname{sech}\!\left(\kappa t+\operatorname{arctanh} z_0\right)\times\nonumber\\
& \times \sqrt{\kappa^2+\left(\frac{B}{\alpha}\right)^2
\tanh^2\!\left(\kappa t+\operatorname{arctanh} z_0\right)} \,.
\end{align}

\section{FitzHugh–Nagumo  neuromorphic map}

In this section we prove that the overdamped Euler-Lagrange-Rayleigh equation for N\'eel vector in antiferromagnetic system  can be reduced locally to an conventional FitzHugh--Nagumo (FHN)
system under an expansion around a suitable operating point. We proceed with Eq.(\ref{solution1}), and introduce the notations to describe dynamics near the north pole, $I_z\simeq 1$:
\begin{eqnarray}\label{eq:Ix-Iz}
    v=I_x, \qquad w=1-I_z.
\end{eqnarray}
Then
\begin{eqnarray}
    I_y=s\sqrt{2w-w^2-v^2}, \qquad s=\pm 1,
    \label{eq:Iy-local}
\end{eqnarray}
where the sign \(s\) selects a local branch of the sphere. The condition for this chart to be valid is
\begin{equation}
    2w-w^2-v^2>0.
\end{equation}
In these variables the exact local equations become
\begin{eqnarray}\label{eq:F-vw}
&&\dot v = F(v,w)=\frac{1-w}{\alpha}\big[-A\,s\sqrt{2w-w^2-v^2}-\nonumber\\
&&\tau p_z v\big],
\end{eqnarray}  
\begin{eqnarray}\label{eq:G-vw}    
&&\dot w = G(v,w)=-\frac{1}{\alpha}\big[C v\,s\sqrt{2w-w^2-v^2}+\nonumber\\
&&\tau p_z(2w-w^2)\big].
\end{eqnarray}
Equations \eqref{eq:F-vw} and \eqref{eq:G-vw} are the correct local two-variable form of the overdamped magnetization dynamics.

The FHN structure is obtained only after a local Taylor expansion and normal-form transformation. Let $(v_\ast,w_\ast)$ be an operating point with
\begin{eqnarray}
q_\ast=2w_\ast-w_\ast^2-v_\ast^2>0,\qquad y_\ast=s\sqrt{q_\ast}\, .
\end{eqnarray}
We introduce two new variables $\xi$ and $\eta$:
\begin{eqnarray}
    v=v_\ast+\xi, \qquad w=w_\ast+\eta.
\end{eqnarray}
After expanding Eq.(\ref{eq:F-vw}) we deduce
\begin{eqnarray}
&&\dot \xi =f_0+f_1\xi+f_2\xi^2+f_3\xi^3+f_w\eta+\nonumber\\
&&\mathcal O(\xi^4,\xi\eta,\eta^2),
    \label{eq:fast-taylor}
\end{eqnarray}
where we introduced the notations
\begin{eqnarray}\label{eq:taylor_f}
&& f_0 = F(v_\ast,w_\ast),\nonumber\\
&& f_1 = F_v(v_\ast,w_\ast),\nonumber\\
&& f_2 = \frac{1}{2}F_{vv}(v_\ast,w_\ast),\\
&& f_3 = \frac{1}{6}F_{vvv}(v_\ast,w_\ast),\nonumber\\
&& f_w = F_w(v_\ast,w_\ast).\nonumber
\end{eqnarray}
and the corresponding derivatives read:
\begin{eqnarray}\label{eq:Fv}
&&F_v =\frac{1-w}{\alpha}\left(A\frac{v}{y}-\tau p_z\right),\nonumber\\
&&F_w =\frac{1}{\alpha}\left[Ay+\tau p_z v-A\frac{(1-w)^2}{y}\right],\\
&&F_{vv}=\frac{A(1-w)}{\alpha}\left(\frac{1}{y}+\frac{v^2}{y^3}\right),\nonumber\\
&&F_{vvv}=\frac{3A(1-w)v}{\alpha}\left(\frac{1}{y^3}+\frac{v^2}{y^5}\right),\nonumber
\end{eqnarray}
with $y=s\sqrt{2w-w^2-v^2}$, evaluated at $(v_\ast,w_\ast)$.
We note that the cubic term in the FHN fast equation does not appear from the lowest-order pole expansion alone.
It appears from the local Taylor expansion of the square-root constraint Eq.(\ref{eq:Iy-local}) around a nonzero
operating point. This requires that $f_3\neq 0$. To eliminate the quadratic term in Eq.(\ref{eq:fast-taylor}), we consider the following shift
\begin{eqnarray}
    \xi=x+h.  \, , \qquad   h=-\frac{f_2}{3f_3} \, . 
\end{eqnarray}
Then the fast FHN equation Eq.(\ref{eq:F-vw}) becomes
\begin{eqnarray}
    \dot x =
    \tilde f_0+\tilde f_1 x+f_3 x^3+f_w\eta
    +\mathcal O(x^4,x\eta,\eta^2),
    \label{eq:fast-shifted}
\end{eqnarray}
where
\begin{eqnarray}
&&\tilde f_1 = f_1-\frac{f_2^2}{3f_3},\\
&&\tilde f_0 =f_0-\frac{f_1f_2}{3f_3}+\frac{2f_2^3}{27f_3^2}.\nonumber
\end{eqnarray}
If
\begin{eqnarray}
    \tilde f_1 f_3 <0,
    \label{eq:cubic-condition}
\end{eqnarray}
then the cubic fast equation can be rescaled to the canonical FHN form. Introducing the rescaled time
\begin{eqnarray}
    \theta=\tilde f_1 t,
\end{eqnarray}
and choosing
\begin{eqnarray}
V=\lambda x, \qquad \lambda^2=-\frac{3f_3}{\tilde f_1},
\end{eqnarray}
we obtain
\begin{eqnarray}\label{eq:FHN-fast-derived}
\frac{dV}{d\theta}=V-\frac{V^3}{3}-W+I_0+\mathcal O(V^4,VW,W^2),   
\end{eqnarray}
where
\begin{eqnarray}\label{eq:I0}
W=-\frac{\lambda f_w}{\tilde f_1}\eta, \qquad I_0=\frac{\lambda \tilde f_0}{\tilde f_1}.
\end{eqnarray}
Thus, the FHN fast equation is a local normal form of the projected N\'eel dynamics. The recovery equation Eq.(\ref{eq:G-vw})   follows from the expansion of $G(v,w)$. To first order,
\begin{eqnarray}
    \dot \eta =
    g_0+g_v \xi+g_w\eta+\mathcal O(\xi^2,\xi\eta,\eta^2),
    \label{eq:recovery-linear}
\end{eqnarray}
where
\begin{eqnarray}\label{1the same for G}
&&g_0 = G(v_\ast,w_\ast),\nonumber\\
&&g_v = G_v(v_\ast,w_\ast),\\
&&g_w = G_w(v_\ast,w_\ast).\nonumber
\end{eqnarray}
Here we introduced the notations:
\begin{eqnarray}\label{2the same for G}
&&G_v(v_\ast,w_\ast)=- \frac{C}{\alpha} \left[ y - \frac{v^2}{y} \right]  ,\\
&&G_w(v_\ast,w_\ast)=-\frac{(1 - w)}{\alpha} \left[ C \frac{v}{y} + 2 \, \tau \, p_z \right].\nonumber
\end{eqnarray}
Using the shift \(\xi=x+h\), we obtain
\begin{eqnarray}
&&\dot\eta =\tilde g_0+g_v x+g_w\eta+\cdots,\\
&&\tilde g_0=g_0+g_v h.\nonumber
\end{eqnarray}
In terms of the normalized variables $V$, $W$, and $\theta$, we obtain
\begin{eqnarray}\label{eq:FHN-recovery-derived}
\frac{dW}{d\theta}=\epsilon(V+a-bW)+\cdots,   
\end{eqnarray}
where we used the notation
\begin{eqnarray}\label{eq:epsilon-general}
&&\epsilon =-\frac{f_w g_v}{\tilde{f}_1^2},\nonumber\\
&& a =\frac{\lambda \tilde g_0}{g_v},\\
&& b =-\frac{g_w}{\epsilon\tilde{f}_1}.\nonumber
\end{eqnarray}
The sign convention is chosen so that $\epsilon>0$. If the product $f_wg_v$ has the opposite sign, the branch choice $s=\pm1$ or the definition of $W$ must be adjusted. 
Thus, finally we obtain AFM neuromorphic equations in the canonical FitzHugh–Nagumo form:
\begin{eqnarray}\label{2FitzHugh–Nagumo}
&&\dot{V}=V-\frac{V^3}{3}-W+I_0,\\
&&\dot{W}=\varepsilon\left(V+a-bW\right).\nonumber
\end{eqnarray}

We note that the current term $I_0$ in Eq.(\ref{2FitzHugh–Nagumo}) is an
effective bias generated by the local normal-form transformation of the magnetic dynamics. Its numerical value and sign depend on the chosen operating point, branch, and normalization convention. An additional externally applied input can be included by replacing
\begin{eqnarray}\label{added 1}
I_0 \longrightarrow I_0+\Delta I_{\rm ext}.
\end{eqnarray}
In a spintronic network, this external term may represent a
spin-pumping signal supplied by an upstream AFM oscillator or by
a transduction circuit.

The equilibrium point for Eq.(\ref{2FitzHugh–Nagumo}) is given by the cubic equation: 
\begin{eqnarray}\label{4FitzHugh–Nagumo}
&&V_0-\frac{V_0^3}{3}+I_0=\frac{V_0+a}{b},\\
&&W_0=\frac{V_0+a}{b}.\nonumber
\end{eqnarray}
The Jacobian matrix
\begin{eqnarray}\label{5FitzHugh–Nagumo}
J=\begin{pmatrix} 1-V_0^2 & -1 \\\\ \varepsilon & -\varepsilon b
\end{pmatrix},
\end{eqnarray}
with $\text{tr}J=\left(1-V_0^2\right)-\varepsilon b$, $\text{det}J=\varepsilon[1-b(1-V_0^2)]$. A Hopf bifurcation occurs when 
\begin{eqnarray}\label{6FitzHugh–Nagumo}
&& \left(1-V_0^2\right)=\varepsilon b,\\
&& \text{det}J>0.\nonumber
\end{eqnarray}
$\text{tr}J=0$, i.e., $1-V_0^2=\varepsilon b$, and $\text{det}J>0$. For fixed values of the reduced parameters $a$, $b$, and $\varepsilon$, the equilibrium branch can be parameterized by the stationary value
$V_0$. From Eq.~(\ref{4FitzHugh–Nagumo}), the corresponding effective bias current is
\begin{equation}
I_0(V_0)
=
\frac{V_0+a}{b}
-
V_0
+
\frac{V_0^3}{3}.
\label{eq:I0_equilibrium_branch}
\end{equation}
The Hopf condition in Eq.~(\ref{6FitzHugh–Nagumo}) then gives the critical stationary
values
\begin{equation}
V_0^{\rm H,\pm}
=
\pm\sqrt{1-\varepsilon b},
\label{eq:Hopf_V0_values}
\end{equation}
provided that $1-\varepsilon b>0$ and $\det J>0$. The corresponding
critical bias currents are
\begin{equation}
I_0^{\rm H,\pm}
=
\frac{V_0^{\rm H,\pm}+a}{b}
-
V_0^{\rm H,\pm}
+
\frac{\left(V_0^{\rm H,\pm}\right)^3}{3}.
\label{eq:Hopf_I0_values}
\end{equation}
The transversality condition is also satisfied away from $V_0=0$,
because
\begin{equation}
\frac{d\,{\rm tr}J}{dI_0}
=
\frac{-2V_0}{b^{-1}-1+V_0^2}
\neq 0
\quad
\text{at}
\quad
V_0=V_0^{\rm H,\pm}.
\label{eq:Hopf_transversality}
\end{equation}
Thus, the effective bias current $I_0$ controls the transition between
a stable fixed point and an oscillatory regime. For the parameter set
used in Case~2 of Table~I, one obtains
$V_0^{\rm H,-}=-0.96438754$ and
$V_0^{\rm H,+}=0.96438754$. The interval between these two Hopf
points corresponds to ${\rm tr}J>0$, so that the stationary state is
unstable and the FHN dynamics develops a self-sustained spiking
cycle.

\section{Results and discussion}

\begin{figure}[htbp]
    \centering
    \includegraphics[width=\columnwidth]{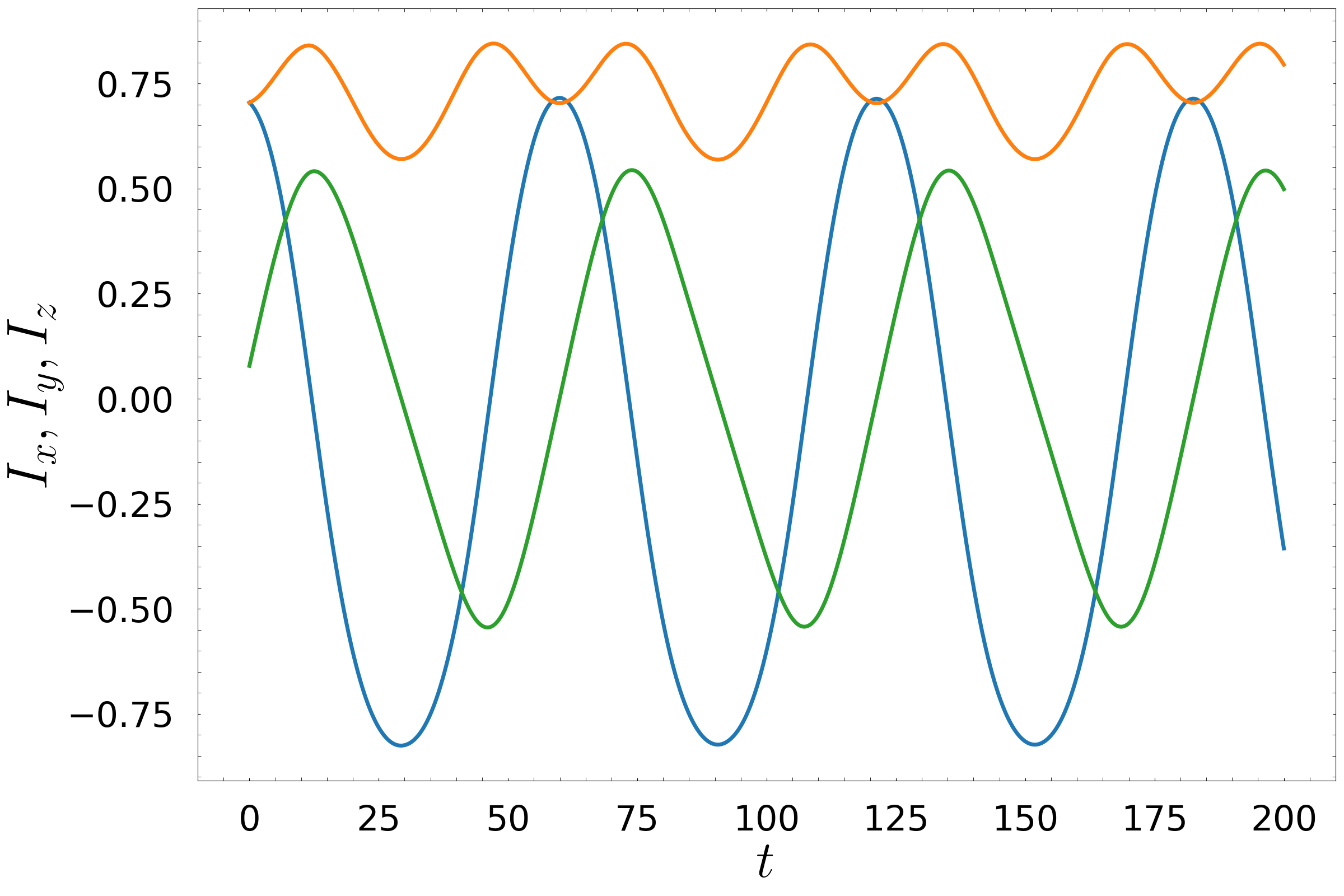}
    \caption{Time dependence of $I_x(t)$(blue), $I_y(t)$(amber), $I_z(t)$ (green) of Eq.(\ref{explicit in components}) with respect to dimensionless time units. The parameters are $\alpha=0.1$, $p_x=0 \, , \, p_z=0.3$, $\tau=0.002$ and $\omega_0=\omega_{ip}=0.1$. The initial conditions are $I_x(0)=I_y(0)=0.7049344$ and $I_z(0)=0.07832604$
    }
    \label{fig:2}
\end{figure}

\begin{figure}[htbp]
    \centering
    \includegraphics[width=\columnwidth]{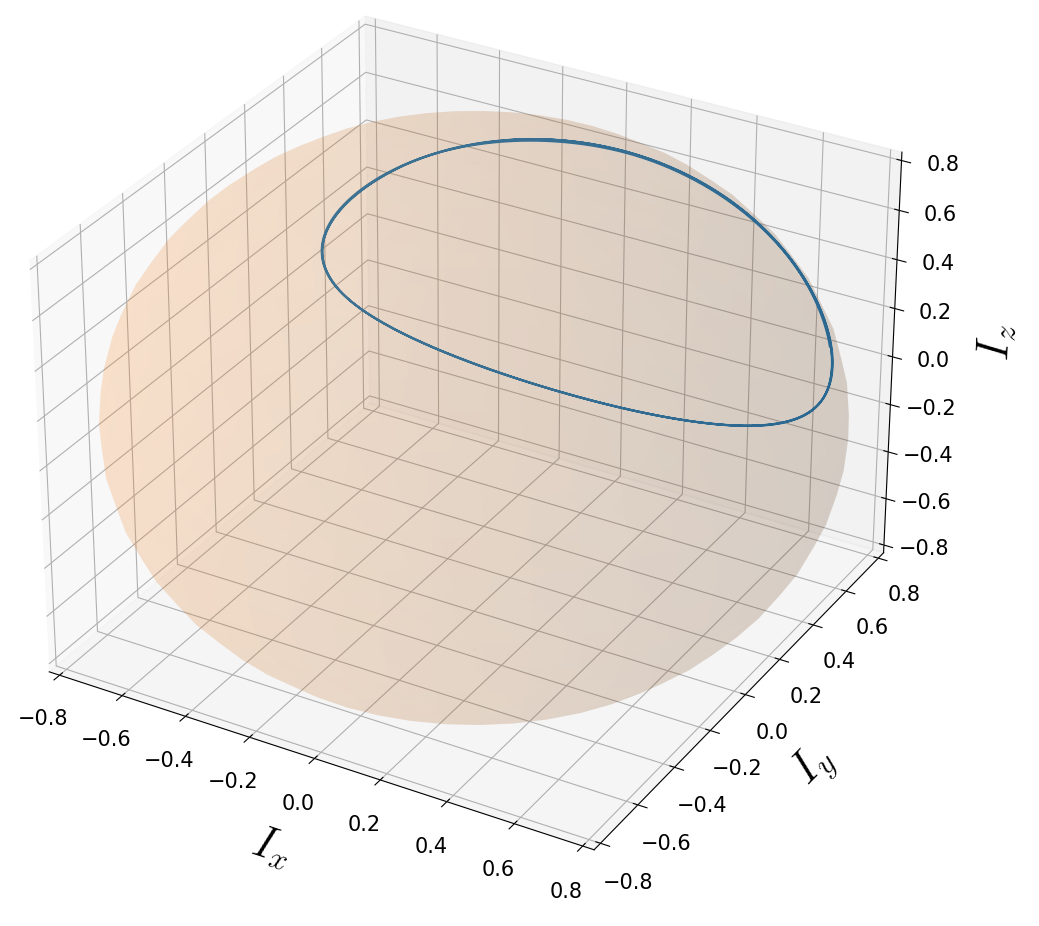}
    \caption{Normalization of the N\'eel vector is preserved. Its norm lies and traces a trajectory on the surface of the Bloch sphere as $I_x(t)$, $I_y(t)$,  $I_z(t)$ evolve in time. The parameters are $\alpha=0.1$, $p_x=0 \, , \, p_z=0.3$, $\tau=0.002$ and $\omega_0=\omega_{ip}=0.1$. The initial conditions are $I_x(0)=I_y(0)=0.7049344$ and $I_z(0)=0.07832604$.}
    \label{fig:3}
\end{figure}

\begin{table*}[ht]
\centering
\renewcommand{\arraystretch}{1.2}
\begin{tabular}{@{}c |c |c| c |c |c |c |c |c@{}}
\hline
Case & $\tau$ & $p_z$ & $v_{\ast}$ & $w_{\ast}$ & $a$ & $b$ & $\epsilon$ & $I_0$ \\
\hline
1 & 4.8000e-02 & 3.1000e-02 & -1.6630e-01 & 3.1193e-01 & 1.2305e+00 & 8.1478e-01 & 8.7228e-02 & -1.5117e+00 \\
2 & 4.8000e-02 & 3.0900e-02 & -1.6624e-01 & 3.1179e-01 & 1.2296e+00 & 8.1969e-01 & 8.5345e-02 & -1.5085e+00 \\
3 & 2.0000e-03 & 9.0000e-01 & -1.6847e-01 & 3.1722e-01 & 1.2894e+00 & 8.8812e-01 & 1.6656e-01 & -1.6956e+00 \\
\hline
\end{tabular}
\medskip
\caption{Values for $v_{\ast}$ and $w_{\ast}$ of local operating point $(v_{\ast},w_{\ast})$ obtained from local Taylor expansion after Eq.(\ref{eq:Ix-Iz}) and Eq.(\ref{eq:Iy-local}) with the corresponding parameters $a$, $b$, $\epsilon$ given by Eq.(\ref{eq:epsilon-general}) and $I_0$ from Eq.(\ref{eq:I0}), for different values of $\tau$ and $p_z$.}
\label{tab:all_cases}
\end{table*}

\begin{table*}[ht]
\centering
\renewcommand{\arraystretch}{1.3}
\begin{tabular}{c |c |c |c |c |c |c |c }
\hline
Case & $v_{\ast}$ & $w_{\ast}$ & $f_0$ & $f_1$ & $f_2$ & $f_3$ &  $f_w$  \\
\hline
1 & -1.6630e-01 & 3.1193e-01 &  -9.5498e-02 & -4.2639e-02 & 1.0282e-01 & -3.4272e-02 & 4.7344e-03 \\
2  & -1.6624e-01 & 3.1179e-01 & -9.5507e-02 & -4.2609e-02 & 1.0285e-01 & -3.4284e-02 & 4.6431e-03  \\
3  & -1.6847e-01 & 3.1722e-01  & -9.5012e-02 & -4.4651e-02 & 1.0143e-01 & -3.3811e-02 & 8.0037e-03 \\
\hline
\end{tabular}
\medskip
\caption{Values for $v_{\ast}$ and $w_{\ast}$ of local operating point $(v_{\ast},w_{\ast})$ obtained from local Taylor expansion after Eq.(\ref{eq:Ix-Iz}) and Eq.(\ref{eq:Iy-local}) with the corresponding values for $f_0$, $f_1$, $f_2$, $f_3$,$f_0$ and $f_w$, from Eq.(\ref{eq:taylor_f}) and Eq.(\ref{eq:Fv}) for the same values of $\tau$ and $p_z$ as in Table~\ref{tab:all_cases}.}
\label{tab:optimized_quantities}
\end{table*}

\begin{table*}[ht]
\centering
\renewcommand{\arraystretch}{1.3}
\begin{tabular}{c |c |c |c |c |c }
\hline
Case & $v_{\ast}$ & $w_{\ast}$  & $g_0$ &  $g_v$ & $g_w$ \\
\hline
1 &  -1.6630e-01 & 3.1193e-01 & 3.9111e-03 & -6.6718e-02 & -4.2768e-03 \\
2 &  -1.6624e-01 & 3.1179e-01  & 3.9332e-03 & -6.6708e-02 & -4.2143e-03 \\
3 &  -1.6847e-01 & 3.1722e-01 & 2.3688e-03 & -6.7101e-02 & -8.3998e-03 \\
\hline
\end{tabular}
\medskip
\caption{Values for $v_{\ast}$ and $w_{\ast}$ of local operating point $(v_{\ast},w_{\ast})$ obtained from local Taylor expansion after Eq.(\ref{eq:Ix-Iz}) and Eq.(\ref{eq:Iy-local}) with the corresponding values for $g_0$, $g_v$ and $g_w$, from Eq.(\ref{1the same for G}) and Eq.(\ref{2the same for G}) for the same values of $\tau$ and $p_z$ as in Table~\ref{tab:all_cases}.}
\label{tab:optimized_quantities_g_values}
\end{table*}

\begin{figure}[htbp]
    \centering
    \includegraphics[width=\columnwidth]{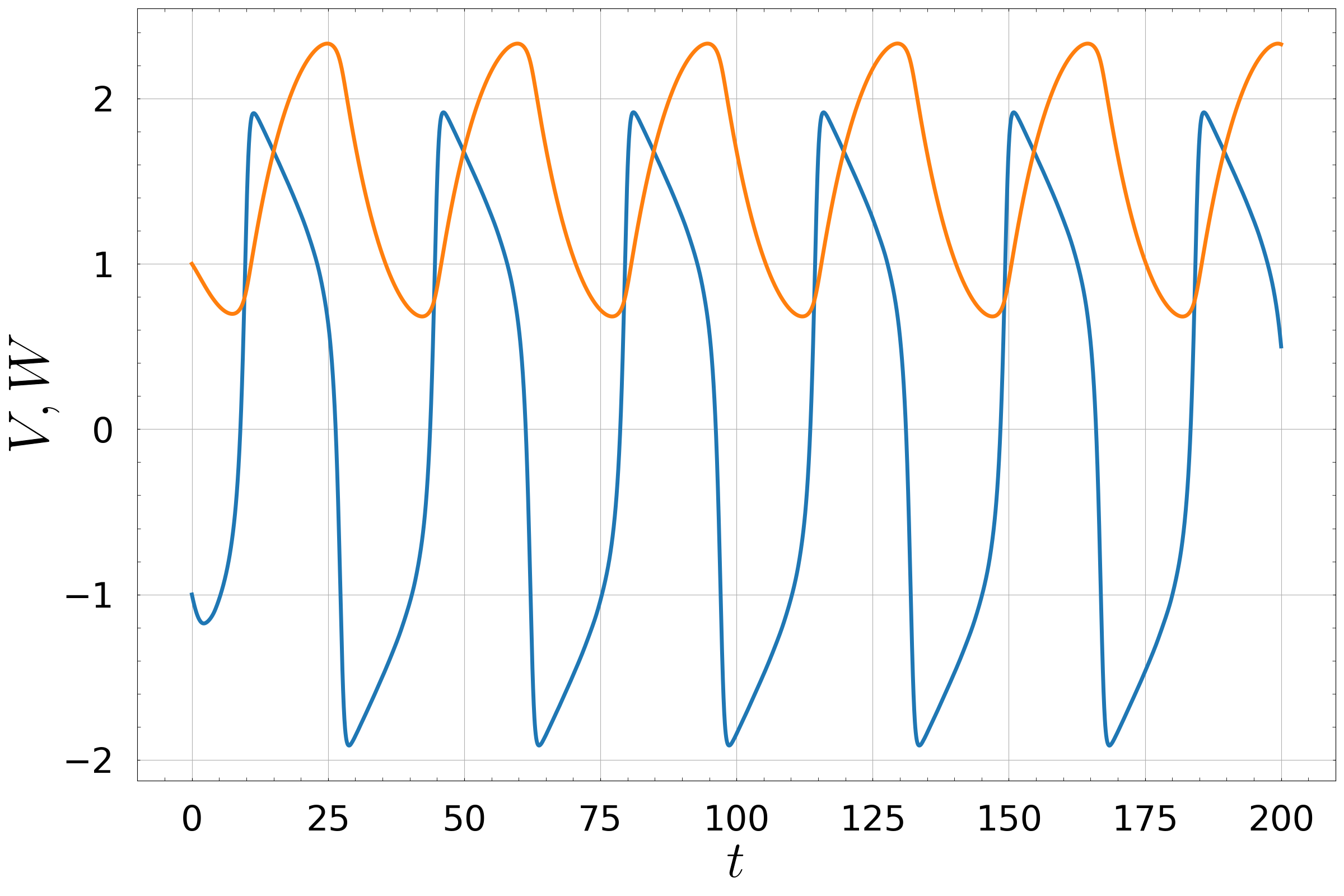}
    \caption{The membrane potential (blue), $V(t)$, with respect to time and the recovery variable (amber), $W(t)$. The parameters are $\alpha=0.1$, $p_z=0.0309$, $\tau=0.048$,  $I_0=1.50848$ and $\omega_0=\omega_{ip}=0.1$. The initial conditions are $V(0)=-1.0$ and $W(0)=1.0$}
    \label{fig:4}
\end{figure}

\newpage
\begin{figure}[htbp]
    \centering
    \includegraphics[width=\columnwidth]{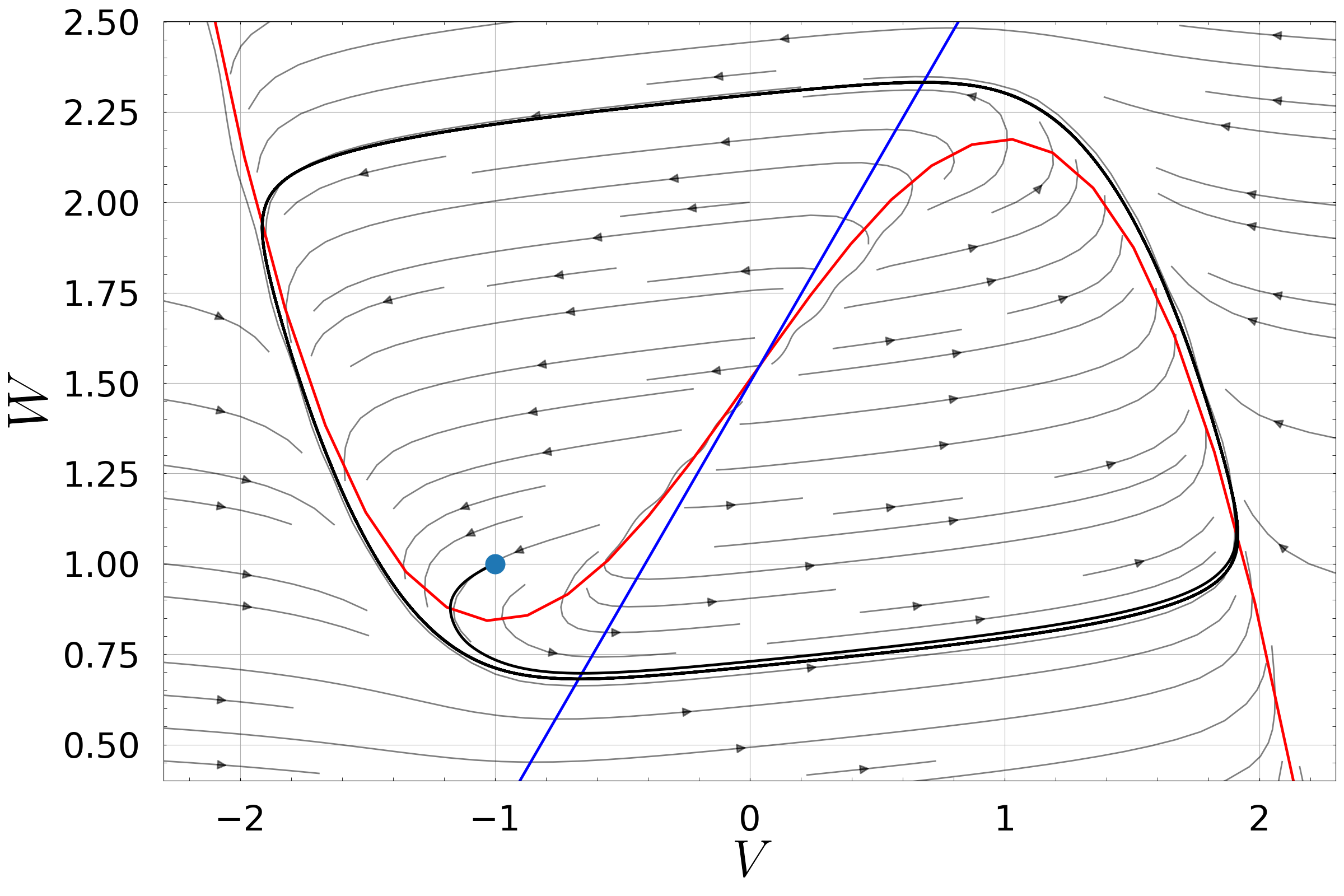}
    \caption{The phase portrait of the trajectory(black), and the nullclines for $\dot{V}=0$ (red) and $\dot{W}=0$ (blue). The parameters are $\alpha=0.1$, $p_z=0.0309$, $\tau=0.048$, $I_0=1.50848$ and $\omega_0=\omega_{ip}=0.1$. The initial conditions are $V(0)=-1.0$ and $W(0)=1.0$}
    \label{fig:5}
\end{figure}

\begin{figure}[htbp]
    \centering
    \includegraphics[width=\columnwidth]{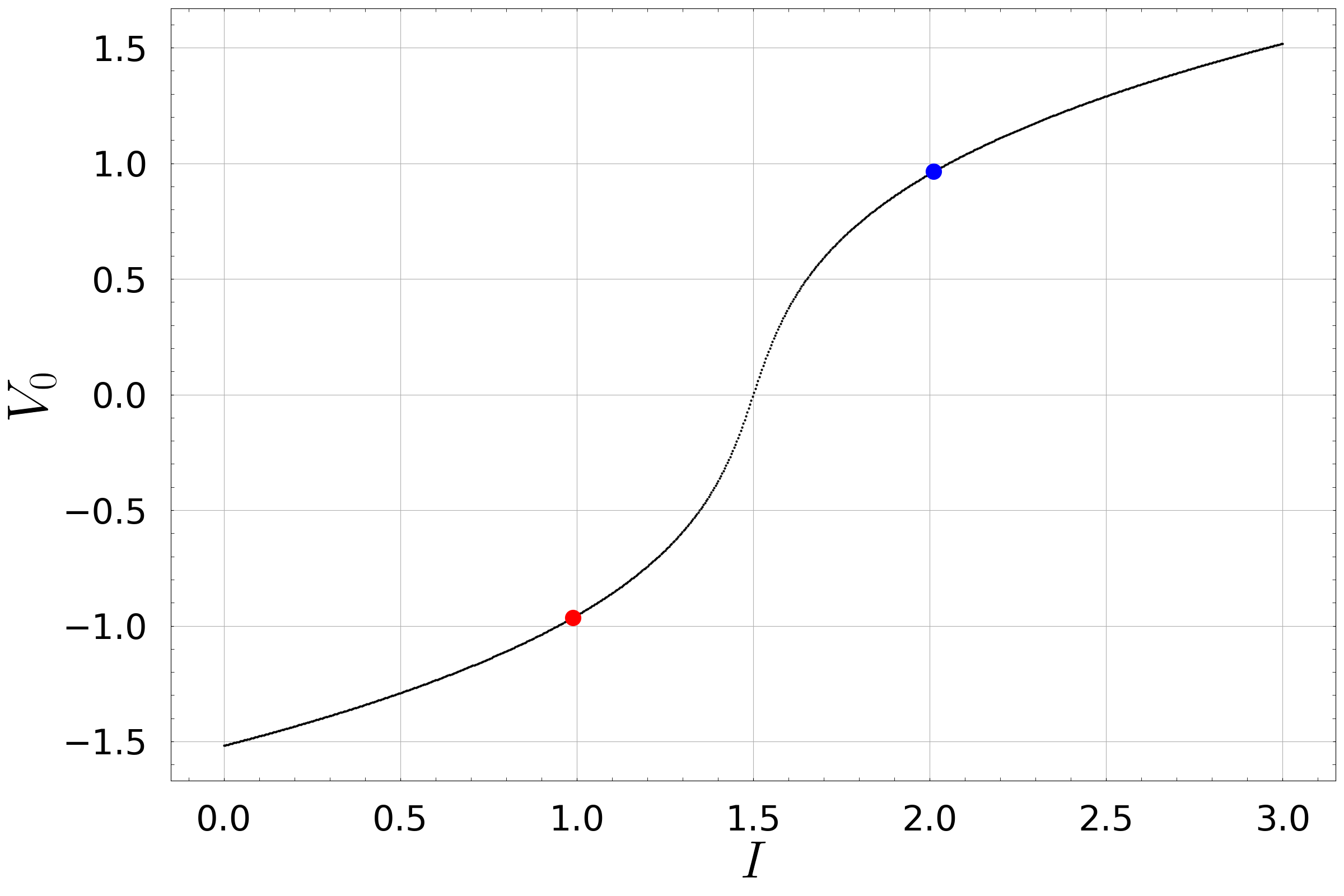}
    \caption{Hopf bifurcation diagram of the AFM-derived FitzHugh--Nagumo
normal form. The gray curve shows the equilibrium branch
$V_0(I_0)$ obtained from Eq.\eqref{eq:I0_equilibrium_branch}.
where $I_0$ is the bifurcation parameter. The red and blue markers denote the
Hopf points satisfying Eq.\eqref{eq:Hopf_V0_values}, located at
$V_0 =-0.96438754$ (red) and
$V_0 =0.96438754$ (blue), respectively.
For the Case~2 parameters of Table~I, the corresponding
critical bias currents are approximately $I_0^{\rm H,-}=0.988956$ and $I_0^{\rm H,+}=2.01119$. The operating value used in Fig.~\ref{fig:4}, $I_0=1.50848$, lies inside this interval. Between the two Hopf points, ${\rm tr}J>0$ and the equilibrium is unstable, giving rise to the oscillatory spiking regime shown in  Fig.~\ref{fig:4}. All remaining parameters correspond to Case~2 of Table~I.}
    \label{fig:6}
\end{figure}

\begin{figure}[htbp]
    \centering
    \includegraphics[width=\columnwidth]{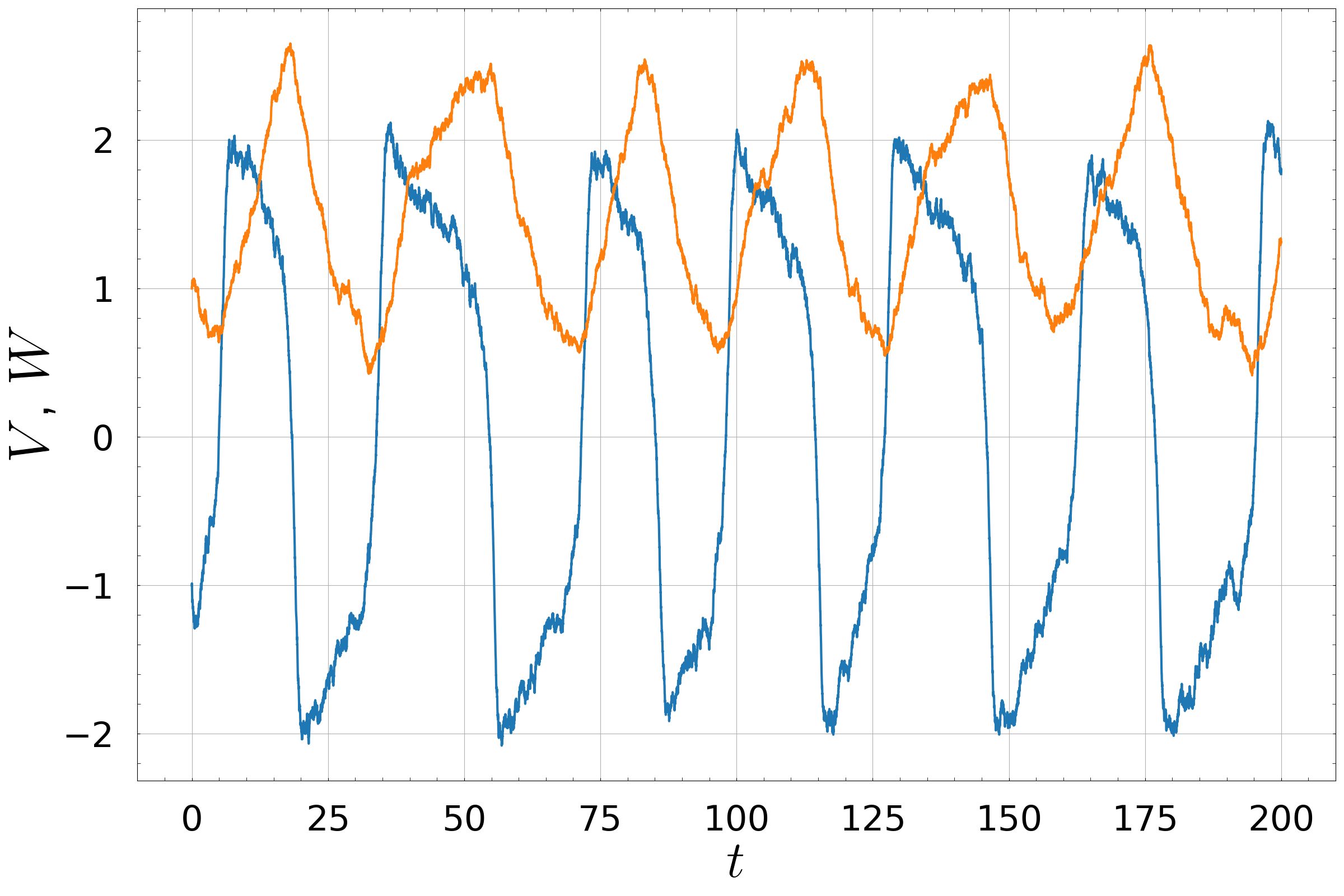}
    \caption{The stochastic membrane potential (blue), $V(t)$, with respect to time and the recovery variable (amber), $W(t)$. The parameters are $\alpha=0.1$, $p_z=0.0309$, $\tau=0.048$,  $I_0=1.50848$ and $\omega_0=\omega_{ip}=0.1$. The initial conditions are $V(0)=-1.0$ and $W(0)=1.0$}
    \label{fig:7}
\end{figure}

\begin{figure}[htbp]
    \centering
    \includegraphics[width=\columnwidth]{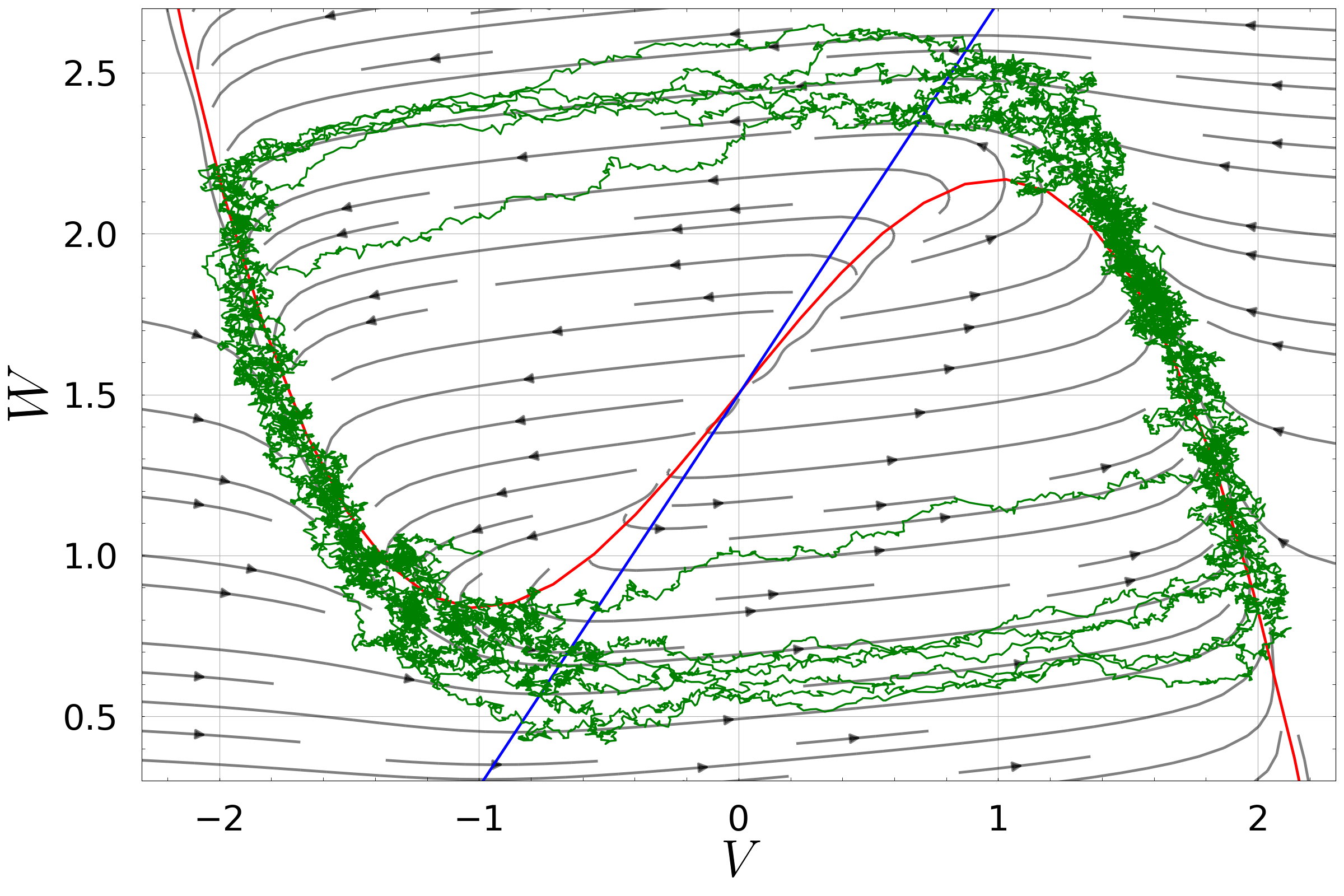}
    \caption{The stochastic phase portrait of the trajectory(green), and the nullclines for $\dot{V}=0$ (red) and $\dot{W}=0$ (blue). The parameters are $\alpha=0.1$, $p_z=0.0309$, $\tau=0.048$, $I_0=1.50848$ and $\omega_0=\omega_{ip}=0.1$. The initial conditions are $V(0)=-1.0$ and $W(0)=1.0$}
    \label{fig:8}
\end{figure}

To finalize the results obtained and conclude the work, we plot the time dependence of projected equations Eq.(\ref{explicit in components}). As we can see, the dynamics is periodic (Fig.~\ref{fig:2}) and the trajectory on the Bloch unit sphere sphere is closed (Fig. \ref{fig:3}). In Fig.~\ref{fig:4} we plot the dynamics of the membrane potential and the recovery variable for the deduced local FitzHugh-Nagumo model Eq.(\ref{2FitzHugh–Nagumo}). We clearly see spikes. In Fig.~\ref{fig:5}, we plot the phase space of the FitzHugh-Nagumo model. The connection between the parameters of the effective local FitzHugh-Nagumo model and the initial neuromorphic system is summarized in Table I.

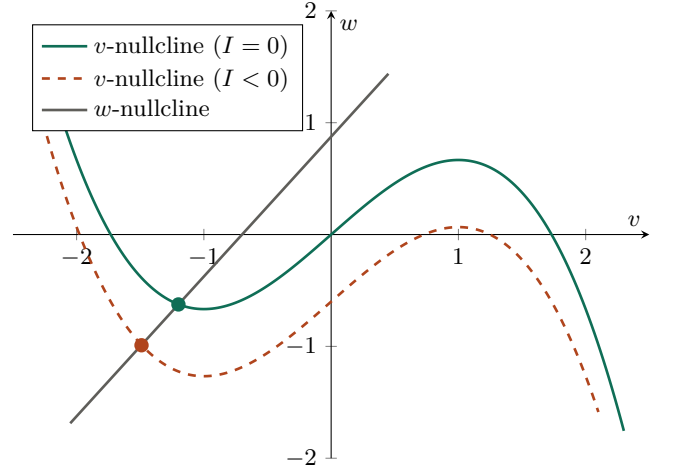
\begin{figure}[h]
\centering
\begin{tikzpicture}
\begin{axis}[
    width=10cm, height=7.5cm,
    xlabel=$v$, ylabel=$w$,
    xmin=-2.5, xmax=2.5, ymin=-2, ymax=2,
    axis lines=middle,
    samples=200, legend pos=north west,
    legend cell align=left, every axis plot/.append style={line width=1pt},
]
\addplot[tealc, domain=-2.3:2.3] {x - (x^3)/3};
\addlegendentry{$v$-nullcline ($I=0$)}
\addplot[coralc, dashed, domain=-2.3:2.1] {x - (x^3)/3 - 0.6};
\addlegendentry{$v$-nullcline ($I<0$)}
\addplot[grayc, domain=-2.05:0.45] {(x + 0.7)/0.8};
\addlegendentry{$w$-nullcline}
\addplot[only marks, mark=*, tealc, mark size=2.2pt]
    coordinates {(-1.2,-0.625)};
\addplot[only marks, mark=*, coralc, mark size=2.2pt]
    coordinates {(-1.49,-0.99)};
\end{axis}
\end{tikzpicture}
\caption{Phase plane of the FitzHugh--Nagumo model. Lowering $I$ slides the
cubic $v$-nullcline down (dashed), carrying the stable rest point down the left
branch to a more hyperpolarized $v$.}
\label{fig:phaseplane}
\end{figure}

The Hopf bifurcation diagram for the membrane potential around the stationary point $V_0$ is shown in Fig.~\ref{fig:6}, using the values $\epsilon$, $a$, and $b$ given in case 2 of Table I. The oscillatory regime of the FitzHugh–Nagumo neuromorphic model lies between the Hopf bifurcation points at $V_0 =-0.96438754$ (red)  and
$V_0 =0.96438754$  (blue). As we already mentioned, the term for the current in our FHN model consists of two terms: $I=I_0+\Delta I_{\rm ext}$, where $I_0$ is the internal term related to the pure magnetization dynamics (effective current) and the second term is the external current term. We note that a positive $I$ raises the cubic (depolarizing); a negative $I$ lowers it leading to the hyperpolarization. Fig.~\ref{fig:phaseplane} shows the standard
FitzHugh parameters $a=0.7$, $b=0.8$, with the resting state at $I=0$ and where it moves under a hyperpolarizing $I<0$.
Physically, a negative $I$ is a \emph{hyperpolarizing} current --- the analogue of injecting anodal (outward) current into a real membrane, the opposite of the depolarizing stimulus that drives the firing. Its effects are as follows: (a) Deeper, more stable resting state. The fixed point slides down the left (attracting) branch of the cubic to a more negative $v$. With the classic parameters, it already sits safely on that branch at $I=0$, so a negative $I$ pushes it \emph{further} onto the stable branch ---the rest point becomes \emph{more} robustly stable, not less. (b)
Suppressed excitability. While the negative current is on, the resting potential is pulled away from the cubic's left knee (the local minimum, the effective firing threshold of the fast subsystem). A given excitatory perturbation now has a larger gap to cross before it can launch the fast jump to the right branch, so the cell is harder to excite. A \emph{steady} negative current also does not produce repetitive firing in the standard regime: the Hopf-bounded oscillatory window for FitzHugh's
parameters lives at positive current, $0.33 \lesssim I \lesssim 1.4$, and lowering $I$ only moves the system away from it. (c) Post-inhibitory rebound (anodal break excitation). This is the genuinely interesting effect. While $I<0$ is held, the slow recovery variable $w$ relaxes down toward the new, lower fixed point. When the current is \emph{switched off} (or stepped back up), the cubic snaps back up but $w$ is momentarily stuck at an anomalously low value. The state then lies below and to
the left of the restored middle branch, and instead of returning quietly to rest it can be flung across into a full large-amplitude excursion --- a spike fired \emph{on release of inhibition}. The FitzHugh approach is used to model and reproduce exactly this phenomenon; it provides a faithful qualitative representation of real anodal-break (rebound) spiking and demonstrates that inhibition can paradoxically trigger a spike.

\section{The stochastic FitzHugh-Nagumo model}

The deterministic FitzHugh--Nagumo equation obtained in Eq.~(\ref{2FitzHugh–Nagumo})
describes the intrinsic local normal form of the overdamped
antiferromagnetic dynamics. In the time variable $\theta$ introduced
in the normal-form reduction, this deterministic model reads
\begin{eqnarray}
&&\frac{dV}{d\theta}=V-\frac{V^{3}}{3}-W+I_{0},\\
&&\frac{dW}{d\theta}=\varepsilon (V+a-bW).\nonumber
\label{eq:det_fhn_theta}
\end{eqnarray}
Here, $V$ is the fast variable, $W$ is the recovery variable, and
$I_{0}$ is the effective bias generated by the local Taylor expansion
and normal-form transformation of the projected N\'eel-vector
dynamics. Thus, $I_{0}$ is an intrinsic parameter of the reduced
magnetic dynamics and should not be identified with the spin-pumping
readout current. For an isolated AFM-FHN neuron, stochasticity can be introduced by
adding thermal, electronic, or interfacial fluctuations directly to the
fast and recovery variables. We therefore write the intrinsic stochastic
AFM-FHN model in the It\^o sense as
\begin{eqnarray}
&&dV_{\theta}=\left[V_{\theta}-\frac{V_{\theta}^{3}}{3}-W_{\theta}+I_{0}\right]d\theta
+\sigma_{V}\,dB_{V}(\theta),\nonumber\\
&&dW_{\theta}=\varepsilon\left(V_{\theta}+a-bW_{\theta}\right)d\theta
+\sigma_{W}\,dB_{W}(\theta),
\label{eq:intrinsic_stochastic_fhn}
\end{eqnarray}
where $B_{V}(\theta)$ and $B_{W}(\theta)$ are independent Brownian
motions, while $\sigma_{V}$ and $\sigma_{W}$ are the noise amplitudes
in the fast and recovery variables, respectively. When
$\sigma_{V}=\sigma_{W}=0$, Eq.~\eqref{eq:intrinsic_stochastic_fhn}
reduces to the deterministic AFM-derived FHN model in Eq.~(\ref{2FitzHugh–Nagumo}). The spin-pumping current generated by the magnetic trajectory is an output of the AFM element. In the present notation, it may be written as follows
\begin{eqnarray}
I_{\rm sp}^{\rm out}(\theta)=\eta g\left|\frac{d\mathbf{I}(\theta)}{d\theta}\right|,
\label{eq:spin_pumping_output}
\end{eqnarray}
up to the same nondimensionalization of time used in the preceding
sections. This signal can be used as a readout current or as a synaptic
signal for another neuron. Therefore, for a single isolated AFM
oscillator, $I_{\rm sp}^{\rm out}$ should not be added back to the
same reduced FHN equation unless an explicit feedback circuit is being
modeled.

If the AFM-FHN neuron is driven by a spin-pumping signal supplied by
an upstream oscillator, a neighboring AFM element, or an external
transduction circuit, the driven stochastic model becomes: 
\begin{align}
&dV_{\theta}=\left[V_{\theta}-\frac{V_{\theta}^{3}}{3}-W_{\theta}+I_{0}
+\chi_{\rm sp} I_{\rm sp}^{\rm ext}(\theta)\right]d\theta + \nonumber\\
&\qquad+\sigma_{V}\,dB_{V}(\theta),\nonumber\\
&dW_{\theta}=\varepsilon\left(V_{\theta}+a-bW_{\theta}\right)d\theta+\sigma_{W}\,dB_{W}(\theta).
\label{eq:driven_stochastic_fhn}
\end{align}
Here, $I_{\rm sp}^{\rm ext}(\theta)$ denotes an externally supplied
spin-pumping input, not the output generated by the same isolated
oscillator. The coefficient $\chi_{\rm sp}$ is the spin-pumping input
coupling strength. With this convention, the effective FHN bias is
shifted according to
\begin{eqnarray}
I_{0}
\longrightarrow
I_{0}+\chi_{\rm sp}I_{\rm sp}^{\rm ext}(\theta).
\label{eq:effective_bias_shift}
\end{eqnarray}

\section{Fokker-Planck equation}

We start derivation of the Fokker-Planck (FP) equation from the driven stochastic FitzHugh--Nagumo system written in It\^o form Eq.(\ref{eq:driven_stochastic_fhn}). We will follow \cite{gardiner2009stochastic,zinn2021quantum}. 
We define the drift vector
\begin{eqnarray}\label{FP1}
\bm{A}(V,W,\theta)
=
\begin{pmatrix}
A_{V}(V,W,\theta)\\[1mm]
A_{W}(V,W)
\end{pmatrix},
\end{eqnarray}
with
\begin{align}
&A_{V}(V,W,\theta)=V-\frac{V^{3}}{3}-W+I_{0}+\chi_{\mathrm{sp}} I_{\mathrm{sp}}^{\mathrm{ext}}(\theta),\nonumber\\
&A_{W}(V,W)=\epsilon(V+a-bW).
\end{align}
The noise matrix is diagonal:
\begin{eqnarray}\label{FP2}
\Sigma=
\begin{pmatrix}
\sigma_{V} & 0\\
0 & \sigma_{W}
\end{pmatrix},
\end{eqnarray}
Therefore the diffusion matrix is
\begin{eqnarray}\label{FP3}
D=
\Sigma\Sigma^{T}=
\begin{pmatrix}
\sigma_{V}^{2} & 0\\
0 & \sigma_{W}^{2}
\end{pmatrix}.
\end{eqnarray}
We introduce a smooth test function of two variables $\phi(V,W)$. Then, according to It\^o's formula:
\begin{eqnarray}
&&d\phi=\frac{\partial \phi}{\partial V}\,dV+\frac{\partial \phi}{\partial W}\,dW+
\frac{1}{2}\frac{\partial^{2}\phi}{\partial V^{2}}(dV)^{2}+\nonumber\\
&&+\frac{1}{2}\frac{\partial^{2}\phi}{\partial W^{2}}(dW)^{2}+\frac{\partial^{2}\phi}{\partial V\partial W}\,dV\,dW .
\end{eqnarray}
Since the Brownian motions are independent, one can write:
\begin{eqnarray}\label{FP4}
&&(dV)^{2}=\sigma_{V}^{2}\,d\theta,\nonumber\\
&&(dW)^{2}=\sigma_{W}^{2}\,d\theta,\\
&&dV\,dW=0.\nonumber
\end{eqnarray}
Therefore,
\begin{align}\label{FP5}
&d\phi=\left[A_{V}\frac{\partial\phi}{\partial V}+A_{W}\frac{\partial\phi}{\partial W}+\frac{\sigma_{V}^{2}}{2}\frac{\partial^{2}\phi}{\partial V^{2}}
+\frac{\sigma_{W}^{2}}{2}\frac{\partial^{2}\phi}{\partial W^{2}}\right]d\theta\nonumber\\
&+\sigma_{V}\frac{\partial\phi}{\partial V}\,dB_{V}+\sigma_{W}\frac{\partial\phi}{\partial W}\,dB_{W}.
\end{align}
Thus, the backward generator reads
\begin{eqnarray}
\mathcal{L}\phi=A_{V}\frac{\partial\phi}{\partial V}+A_{W}\frac{\partial\phi}{\partial W}
+\frac{\sigma_{V}^{2}}{2}\frac{\partial^{2}\phi}{\partial V^{2}}+\frac{\sigma_{W}^{2}}{2}\frac{\partial^{2}\phi}{\partial W^{2}}.
\end{eqnarray}
Taking the expectation value gives the following
\begin{eqnarray}\label{FP6}
\frac{d}{d\theta}\mathbb{E}\left[\phi(V_{\theta},W_{\theta})\right]=\mathbb{E}
\left[\mathcal{L}\phi(V_{\theta},W_{\theta})\right].
\end{eqnarray}
If $P(V,W,\theta)$ is the probability density of the random variables
$(V_{\theta},W_{\theta})$, then
\begin{eqnarray}\label{FP7}
\mathbb{E}\left[\phi(V_{\theta},W_{\theta})\right]=\int_{\mathbb{R}^{2}}\phi(V,W)P(V,W,\theta)\,dV\,dW.
\end{eqnarray}
Hence,
\begin{eqnarray}\label{FP8}
\frac{d}{d\theta}\int_{\mathbb{R}^{2}}\phi P\,dV\,dW=\int_{\mathbb{R}^{2}}(\mathcal{L}\phi)P\,dV\,dW.
\end{eqnarray}
By integration by parts, assuming that \(P\) and its derivatives decay
sufficiently fast at infinity, the derivatives can be transferred from
\(\phi\) onto \(P\). This gives the adjoint equation
\begin{eqnarray}\label{FP9}
\frac{\partial P}{\partial \theta}=\mathcal{L}^{*}P.
\end{eqnarray}
The adjoint operator is
\begin{eqnarray}\label{FP9a}
&&\mathcal{L}^{*}P=-\frac{\partial}{\partial V}\left(A_{V}P\right)-\frac{\partial}{\partial W}\left(A_{W}P\right)+
\frac{\sigma_{V}^{2}}{2}\frac{\partial^{2}P}{\partial V^{2}}\nonumber\\
&&+\frac{\sigma_{W}^{2}}{2}
\frac{\partial^{2}P}{\partial W^{2}}.
\end{eqnarray}
Substituting $A_{V}$ and $A_{W}$, we obtain FP equation for the probability density
$P(V,W,\theta)$:
\begin{eqnarray}
&&\frac{\partial P}{\partial \theta}=-\frac{\partial}{\partial V}
\left\{\left[V-\frac{V^{3}}{3}-W+I_{0}+\chi_{\rm sp}I_{\rm sp}^{\rm ext}(\theta)
\right]P\right\}\nonumber\\
&&-\frac{\partial}{\partial W}\left\{\varepsilon(V+a-bW)P
\right\}+\frac{\sigma_{V}^{2}}{2}\frac{\partial^{2}P}{\partial V^{2}}+\\
&&+\frac{\sigma_{W}^{2}}{2}\frac{\partial^{2}P}{\partial W^{2}} .\nonumber
\label{eq:fokker_planck_driven}
\end{eqnarray}

\section{Analytical solution of FP equation}

When $I_{\mathrm{sp}}^{\mathrm{ext}}(\theta)=0$, i.e, no external current, we have only internal effective current, the stochastic system slightly simplifies: 
\begin{eqnarray}\label{FP11}
&&dV_{\theta}=\left[V_{\theta}-\frac{V_{\theta}^{3}}{3}-W_{\theta}+I_{0}\right]d\theta+\sigma_{V}\,dB_{V}(\theta),\\
&&dW_{\theta}=\epsilon\left(V_{\theta}+a-bW_{\theta}\right)d\theta+\sigma_{W}\,dB_{W}(\theta).\nonumber
\end{eqnarray}
The corresponding Fokker--Planck equation is
\begin{eqnarray}\label{FP12}
&&\frac{\partial P}{\partial \theta}=-\frac{\partial}{\partial V}\left[\left(V-\frac{V^{3}}{3}-W+I_{0}\right)P\right]-\\
&&-\frac{\partial}{\partial W}\left[\epsilon(V+a-bW)P\right]+\frac{\sigma_{V}^{2}}{2}\frac{\partial^{2}P}{\partial V^{2}}+\frac{\sigma_{W}^{2}}{2}\frac{\partial^{2}P}{\partial W^{2}}.\nonumber
\end{eqnarray}
For convenience, define
\begin{eqnarray}\label{FP13}
&&f(V,W)=V-\frac{V^{3}}{3}-W+I_{0},\nonumber\\
&&g(V,W)=\epsilon(V+a-bW),\nonumber\\
&&D_{V}=\frac{\sigma_{V}^{2}}{2},\\
&&D_{W}=\frac{\sigma_{W}^{2}}{2}.\nonumber
\end{eqnarray}
Then the equation can be written compactly as
\begin{eqnarray}\label{FP14}
&&\frac{\partial P}{\partial \theta}=-\partial_{V}(fP)-\partial_{W}(gP)+D_{V}\partial_{V}^{2}P+\\
&& + D_{W}\partial_{W}^{2}P.\nonumber
\end{eqnarray}
Equivalently, in probability-current form:
\begin{eqnarray}\label{FP15}
&&\frac{\partial P}{\partial \theta}=-\partial_{V}J_{V}-\partial_{W}J_{W},
\end{eqnarray}
where
\begin{eqnarray}\label{FP16}
&&J_{V}=fP-D_{V}\partial_{V}P,\\
&&J_{W}=gP-D_{W}\partial_{W}P.\nonumber
\end{eqnarray}
Let
\begin{eqnarray}\label{FP17}
\mathcal{L}_{0}^{*}=-\partial_{V}(f\,\cdot)-\partial_{W}(g\,\cdot)+D_{V}\partial_{V}^{2}+D_{W}\partial_{W}^{2}.
\end{eqnarray}
For an initial probability density
\begin{eqnarray}\label{FP18}
P(V,W,0)=P_{0}(V,W),
\end{eqnarray}
the exact formal solution is 
\begin{eqnarray}
P(V,W,\theta)=\left[e^{\theta\mathcal{L}_{0}^{*}}P_{0}\right](V,W).
\end{eqnarray}
Equivalently, if the initial condition is a point source 
\begin{eqnarray}\label{FP19}
P(V,W,0)=\delta(V-V_{i})\delta(W-W_{i}).
\end{eqnarray}
Then
\begin{eqnarray}\label{FP20}
\frac{\partial G_{\theta}}{\partial \theta}=\mathcal{L}_{0}^{*}G_{\theta},
\end{eqnarray}
where the Green function satisfies
\begin{eqnarray}\label{FP21}
P(V,W,\theta)=G_{\theta}(V,W\mid V_{i},W_{i}),
\end{eqnarray}
and
\begin{eqnarray}\label{FP22}
G_{0}(V,W\mid V_{i},W_{i})=\delta(V-V_{i})\delta(W-W_{i}).
\end{eqnarray}
The solution for a general initial density is then
\begin{align}\label{FP23}
& P(V,W,\theta)= \nonumber \\
& = \int_{\mathbb{R}^{2}}G_{\theta}(V,W\mid V',W')P_{0}(V',W')\,dV'\,dW'.
\end{align}
A discrete time-sliced representation of the Green function reads
\begin{align}\label{FP24}
& G_{\theta}(x_{N}\mid x_{0})=\lim_{N\to\infty} \left[ \int\prod_{n=1}^{N-1}dV_{n}\,dW_{n} \times \right. \nonumber \\ 
& \left. \times \prod_{n=0}^{N-1}K_{\Delta\theta}(x_{n+1}\mid x_{n}) \right] ,
\end{align}
where
\begin{eqnarray}\label{FP25}
x_{n}=(V_{n},W_{n}),\qquad \Delta\theta=\frac{\theta}{N}.
\end{eqnarray}
The short-time kernel reads
\begin{align}\label{FP26}
& K_{\Delta\theta}(x_{n+1}\mid x_{n})  =\frac{1}{2\pi\sigma_{V}\sigma_{W}\Delta\theta}\times \nonumber \\
& \times \exp\left[
-\frac{\left(V_{n+1}-V_{n}-f(V_{n},W_{n})\Delta\theta\right)^{2}}{2\sigma_{V}^{2}\Delta\theta} - \right. \nonumber \\
& \left. -\frac{\left(W_{n+1}-W_{n}-g(V_{n},W_{n})\Delta\theta\right)^{2}}{2\sigma_{W}^{2}\Delta\theta}\right].
\end{align}
Because the drift contains the nonlinear cubic term $-V^{3}/3$, the full nonlinear Fokker--Planck equation does not reduce to a closed elementary Gaussian density.
Moreover, the stationary solution is not generally a zero-current Boltzmann distribution. A zero-current stationary state would require
\begin{eqnarray}\label{FP27}
J_{V}=0,\qquad J_{W}=0.
\end{eqnarray}
That is,
\begin{eqnarray}\label{FP28}
&&D_{V}\partial_{V}P_{\infty}=fP_{\infty},\\
&&D_{W}\partial_{W}P_{\infty}=gP_{\infty}.\nonumber
\end{eqnarray}
Equivalently,
\begin{eqnarray}\label{FP29}
&&\partial_{V}\ln P_{\infty}=\frac{f}{D_{V}},\\
&&\partial_{W}\ln P_{\infty}=\frac{g}{D_{W}}.\nonumber
\end{eqnarray}
For such a potential form to exist, the compatibility condition must hold:
\begin{eqnarray}
\partial_{W}\left(\frac{f}{D_{V}}\right)=\partial_{V}\left(\frac{g}{D_{W}}\right).
\end{eqnarray}
Taking into account 
\begin{eqnarray}\label{FP30}
&&\partial_{W}\left(\frac{f}{D_{V}}\right)=-\frac{1}{D_{V}},\\
&&\partial_{V}\left(\frac{g}{D_{W}}\right)=\frac{\epsilon}{D_{W}}.\nonumber
\end{eqnarray}
Therefore, for ${V}>0$, $D_{W}>0$, and $\epsilon>0$, we have
\begin{eqnarray}\label{FP31}
-\frac{1}{D_{V}}\neq\frac{\epsilon}{D_{W}}.
\end{eqnarray}
Thus, the stationary state, when it exists, has a nonzero circulating
probability current.

While the full nonlinear equation does not have a simple closed elementary solution, a closed Gaussian solution is obtained by linearizing the dynamics near a deterministic fixed point.
The deterministic fixed point satisfies
\begin{eqnarray}\label{FP32}
&&V_{0}-\frac{V_{0}^{3}}{3}+I_{0}=W_{0},\\
&&W_{0}=\frac{V_{0}+a}{b}.\nonumber
\end{eqnarray}
Equivalently,
\begin{eqnarray}\label{FP33}
V_{0}-\frac{V_{0}^{3}}{3}+I_{0}=\frac{V_{0}+a}{b}.
\end{eqnarray}
Introduce local variables
\begin{eqnarray}\label{FP34}
u=V-V_{0},\qquad r=W-W_{0},\qquad x=
\begin{pmatrix}
u\\ r
\end{pmatrix}.
\end{eqnarray}
Linearizing the drift gives the Ornstein--Uhlenbeck system
\begin{eqnarray}\label{FP35}
dx=Jx\,d\theta+\Sigma\,dB,
\end{eqnarray}
where
\begin{eqnarray}\label{FP36}
J=
\begin{pmatrix}
1-V_{0}^{2} & -1\\
\epsilon & -\epsilon b
\end{pmatrix},
\qquad
\Sigma
=
\begin{pmatrix}
\sigma_{V} & 0\\
0 & \sigma_{W}
\end{pmatrix}.
\end{eqnarray}
Let
\begin{eqnarray}\label{FP37}
Q=\Sigma\Sigma^{T}
=
\begin{pmatrix}
\sigma_{V}^{2} & 0\\
0 & \sigma_{W}^{2}
\end{pmatrix}.
\end{eqnarray}
The transition probability density of the linearized problem is exactly
Gaussian:
\begin{align}\label{FP38}
& p(x,\theta\mid x_{0})=\frac{1}{2\pi\sqrt{\det C(\theta)}} \, \times \nonumber \\
& \times \exp\left[-\frac{1}{2}\left(x-e^{J\theta}x_{0}
\right)^{T}C(\theta)^{-1}\left(x-e^{J\theta}x_{0}\right)\right],
\end{align}
where
\begin{eqnarray}\label{FP39}
C(\theta)=\int_{0}^{\theta}e^{Js}Qe^{J^{T}s}\,ds.
\end{eqnarray}
The matrix exponential can be written explicitly as
\begin{eqnarray}\label{FP40}
&&e^{J\theta}=e^{\frac{\operatorname{tr}J}{2}\theta}\left[\cosh(\Omega\theta)I+
\frac{\sinh(\Omega\theta)}{\Omega}\left(J-\frac{\operatorname{tr}J}{2}I\right)\right],\nonumber\\
&&\Omega=\sqrt{\frac{(\operatorname{tr}J)^{2}}{4}-\det J}.
\end{eqnarray}
If the fixed point is stable, namely
\begin{eqnarray}\label{FP41}
\operatorname{tr}J<0,\qquad\det J>0,
\end{eqnarray}
then the long-time stationary density of the linearized problem is
\begin{eqnarray}\label{FP43}
p_{\infty}(x)=\frac{1}{2\pi\sqrt{\det \Sigma_{\infty}}}\exp\left[-\frac{1}{2}x^{T}\Sigma_{\infty}^{-1}x\right],
\end{eqnarray}
where $\Sigma_{\infty}$ solves the Lyapunov equation
\begin{eqnarray}\label{FP44}
J\Sigma_{\infty}+\Sigma_{\infty}J^{T}+Q=0.
\end{eqnarray}
Considering 
\begin{eqnarray}
c=1-V_{0}^{2},\qquad q_{V}=\sigma_{V}^{2},\qquad q_{W}=\sigma_{W}^{2},
\end{eqnarray}
and
\begin{eqnarray}\label{FP45}
\Sigma_{\infty}=
\begin{pmatrix}
S_{11} & S_{12}\\
S_{12} & S_{22}
\end{pmatrix},
\end{eqnarray}
one obtains
\begin{eqnarray}\label{FP46}
S_{11}=\frac{\epsilon q_{V}(1-bc+b^{2}\epsilon)+q_{W}}{2\epsilon(1-bc)(\epsilon b-c)}.
\end{eqnarray}
Furthermore,
\begin{eqnarray}\label{FP47}
S_{12}=cS_{11}+\frac{q_{V}}{2},
\end{eqnarray}
and
\begin{eqnarray}\label{FP48}
S_{22}=\frac{1}{b}\left(S_{12}+\frac{q_{W}}{2\epsilon}\right).
\end{eqnarray}

For numerical integration, we use the Euler--Maruyama method. With time
step $\Delta \theta$ and independent standard normal random variables
$\zeta_{V,n}$ and $\zeta_{W,n}$, Eq.~\eqref{eq:driven_stochastic_fhn}
is discretized as
\begin{eqnarray}
&&V_{n+1}=V_{n}+\left[V_{n}-\frac{V_{n}^{3}}{3}-W_{n}+I_{0}
+\chi_{\rm sp}I_{\rm sp}^{\rm ext}(\theta_{n})\right]\Delta\theta\nonumber\\
&&+\sigma_{V}\sqrt{\Delta\theta}\,\zeta_{V,n},\\
&&W_{n+1}=W_{n}+\varepsilon(V_{n}+a-bW_{n})\Delta\theta
+\sigma_{W}\sqrt{\Delta\theta}\,\zeta_{W,n}.\nonumber
\label{eq:euler_maruyama_driven}
\end{eqnarray}
In the absence of an externally supplied spin-pumping input,
$I_{\rm sp}^{\rm ext}=0$, this scheme gives the isolated stochastic
AFM-FHN neuron. In the deterministic limit,
$\sigma_{V}=\sigma_{W}=0$ and $I_{\rm sp}^{\rm ext}=0$, it reduces to
 Eq.~(\ref{2FitzHugh–Nagumo}).

Figure~\ref{fig:7}, shows the stochastic time traces of $V(t)$ (blue) and $W(t)$ (amber). The fast variable retains relaxation-oscillation spikes, but the spike times and peak values fluctuate from cycle to cycle. Figure~\ref{fig:8}, presents the corresponding enlarged stochastic phase portrait. The red and blue curves are the deterministic nullclines $\dot V=0$ and $\dot W=0$, and the vector field is the deterministic drift. The green trajectory follows the same slow-fast loop as the deterministic orbit but is broadened by noise. Thus, the stochastic AFM-derived FHN neuron remains organized by the deterministic nullcline geometry while acquiring realistic variability in threshold crossing, spike timing, and recovery. This behavior is useful for neuromorphic applications because controlled noise can support probabilistic spiking, stochastic resonance, and robust signal detection in subthreshold regimes.

\section{Conclusions}

We have derived an overdamped, sphere-constrained dynamical model for the N\'eel vector of an antiferromagnetic spin-Hall nano-oscillator and shown that it preserves the normalization of the order parameter. For spin polarization along the easy axis, the dynamics reduces to an asymmetric rotator, allowing analytic solutions in selected limits and a closed expression for the spin-pumping current. Near a nonzero local operating point, the projected AFM equations can be transformed into a FitzHugh-Nagumo normal form. This mapping identifies the effective membrane variable, recovery variable, bias current, and Hopf condition directly in terms of magnetic and spin-torque parameters.

The stochastic extension demonstrates how thermal and electronic fluctuations can be incorporated into the intrinsic AFM-derived FHN neuron, while spin-pumping signals can enter as externally
supplied inputs in a driven or network setting. The enlarged stochastic phase portrait shows that the deterministic nullclines continue to organize the noisy trajectory, while noise provides spike-to-spike variability. These results connect AFM material parameters with canonical neuromorphic dynamics and provide a compact analytic platform for designing ultrafast AFM spiking elements. Future work can extend the present single-neuron reduction to coupled AFM oscillator arrays, experimentally calibrated noise amplitudes, and learning rules based on spin-pumping-mediated synaptic currents.

\textbf{Acknowledgements} The author D. Maroulakos gratefully acknowledges support from the Doctoral School at the University of Rzeszów and the NAWA STER scholarship program for the best foreign doctoral students studying at the Doctoral School at the University of Rzesz\'ow within the framework of the Project ``UR an international PhD student'' (Decision of the Director of NAWA No. BPI/STE/2023/1/00001/DEC/01 dated 19.10.2023) in accordance with the assumptions of the NAWA STER Program - Internationalization of Doctoral Schools.

\appendix

\section{FitzHugh–Nagumo  neuromorphic map}

For the overdamped system, we obtained the set of the projected equations:

\begin{eqnarray}\nonumber
&&\alpha\dot I_x=-\left(\omega_0^2+\omega_{ip}^2\right) I_y I_z + \\ \nonumber
&& +\tau\left[ p_x-(p_x I_x+p_z I_z)I_x\right] ,\nonumber\\
&&\alpha\dot I_y=\,\omega_0^2\, I_x I_z-\tau\,(p_x I_x+p_z I_z)\,I_y,\\
&&\alpha\dot I_z=\omega_{ip}^2\, I_x I_y+\tau\left[ p_z-(p_x I_x+p_z I_z)I_z\right].\nonumber
\end{eqnarray}
The derivation of the local FitzHugh-Nagumo (FHN) model requires several steps: 
We introduce the notations to describe dynamics near the north pole, $I_z\simeq 1$:
\begin{eqnarray}
    v=I_x, \qquad w=1-I_z.
\end{eqnarray}
Then
\begin{eqnarray}
    I_y=s\sqrt{2w-w^2-v^2}, \qquad s=\pm 1,
    \label{eq:Iy-local-appendix}
\end{eqnarray}
where the sign \(s\) selects a local branch of the sphere. The condition for this chart to be valid is
\begin{equation}
    2w-w^2-v^2>0.
\end{equation}

Finally, we get:
\begin{eqnarray}\label{appendixFHN:F-vw}
&&\dot v = F(v,w)=\frac{1-w}{\alpha}\big[-A\,s\sqrt{2w-w^2-v^2}-\nonumber\\
&&\tau p_z v\big],
\end{eqnarray}  
\begin{eqnarray}\label{appendixFHN:G-vw}    
&&\dot w = G(v,w)=-\frac{1}{\alpha}\big[C v\,s\sqrt{2w-w^2-v^2}+\nonumber\\
&&\tau p_z(2w-w^2)\big].
\end{eqnarray}

The FHN structure is obtained only after a local Taylor expansion and normal-form transformation. Let $(v_\ast,w_\ast)$ be an operating point with
\begin{eqnarray}
q_\ast=2w_\ast-w_\ast^2-v_\ast^2>0,\qquad y_\ast=s\sqrt{q_\ast}\, .
\end{eqnarray}
We introduce two new variables $\xi$ and $\eta$:
\begin{eqnarray}
    v=v_\ast+\xi, \qquad w=w_\ast+\eta.
\end{eqnarray}
The Taylor expansion is then the following
 \begin{eqnarray}\label{eq:taylor_appendix}\nonumber
&& \frac{\mathrm{d}}{\mathrm{d} t}(v_\ast+\xi) = \dot{\xi} =  F(v,w) =  F(v_\ast,w_\ast) + \\ \nonumber
&& + \frac{\partial F}{\partial v}\bigg|_{v=v_\ast} (v - v_\ast) + \frac{\partial^2 F}{\partial v^2}\bigg|_{v=v_\ast} \frac{(v - v_\ast)^2}{2} +\\ \nonumber
&& +  \frac{\partial^3 F}{\partial v^3}\bigg|_{v=v_\ast} \frac{(v - v_\ast)^3}{6} +\cdots\\ \nonumber
&& + \frac{\partial F}{\partial w}\bigg|_{w=w_\ast} (w - w_\ast) + \cdots \\ 
&& +  \frac{\partial^2 F}{\partial v \partial w}\bigg|_{\substack{v=v_\ast\\w=w_{\ast}}} \frac{(v - v_\ast) (w - w_\ast)}{2} + \cdots\\ \nonumber
&&=  F(v_\ast,w_\ast) + F_v (v_\ast,w_\ast) \xi + F_{vv} (v_\ast,w_\ast) \frac{1}{2}\xi^2 + \\ \nonumber 
&& +  F_{vvv} (v_\ast,w_\ast) \frac{1}{6}\xi^3 +\cdots\\ \nonumber
&& + F_{w} (v_\ast,w_\ast) \eta + \cdots \\ \nonumber
&& +  F_{vw} (v_\ast,w_\ast) \frac{1}{2}\xi \eta + \cdots
\end{eqnarray}
Using the notations:
\begin{eqnarray}\label{eq:f0_f1_f2_f3_fw}
&& f_0 = F(v_\ast,w_\ast),\nonumber\\
&& f_1 = F_v(v_\ast,w_\ast),\nonumber\\
&& f_2 = \frac{1}{2}F_{vv}(v_\ast,w_\ast),\\
&& f_3 = \frac{1}{6}F_{vvv}(v_\ast,w_\ast),\nonumber\\
&& f_w = F_w(v_\ast,w_\ast).\nonumber
\end{eqnarray}
the corresponding derivatives read:
\begin{eqnarray}
&&F_v =\frac{1-w}{\alpha}\left(A\frac{v}{y}-\tau p_z\right),\nonumber\\
&&F_w =\frac{1}{\alpha}\left[A y + \tau \, p_z \, v - A \frac{(1-w)^2}{y} \right], \\
&&F_{vv}=\frac{A(1-w)}{\alpha}\left(\frac{1}{y}+\frac{v^2}{y^3}\right),\nonumber\\
&&F_{vvv}=\frac{3A(1-w)v}{\alpha}\left(\frac{1}{y^3}+\frac{v^2}{y^5}\right),\nonumber
\end{eqnarray}
with $y=s\sqrt{2w-w^2-v^2}$, evaluated at $(v_\ast,w_\ast)$.
We consider the following shift
\begin{eqnarray}\label{eq:h_definition}
    \xi=x+h, \qquad   h=-\frac{f_2}{3f_3} \, ,
\end{eqnarray}
the fast FHN,~Eq.(\ref{appendixFHN:F-vw}) using the Taylor expansion,~Eq.(\ref{eq:taylor_appendix}) and the notation in~Eq.(\ref{eq:f0_f1_f2_f3_fw}) becomes
\begin{eqnarray} \nonumber
\dot{x} =&&  f_0 + f_1 (x + h) + f_2 (x + h)^2  +  \\ \nonumber
&& + f_3 (x+h)^3 + f_w \, \eta + \ldots \, .\\ \nonumber 
\end{eqnarray}
Expanding the terms in brackets and using $h=-\frac{f_2}{3f_3}$, Eq.(\ref{eq:h_definition}), to eliminate the quadratic term, after some algebra, we finally get
\begin{eqnarray}
    \dot x = \tilde f_0+\tilde f_1 x+f_3 x^3+f_w\eta+\mathcal O(x^4,x\eta,\eta^2),
    \label{appendixFHN:fast-shifted}
\end{eqnarray}
where
\begin{eqnarray}
&&\tilde f_1 = f_1-\frac{f_2^2}{3f_3},\\
&&\tilde f_0 =f_0-\frac{f_1f_2}{3f_3}+\frac{2f_2^3}{27f_3^2}.\nonumber
\end{eqnarray}
If the condition
\begin{eqnarray}
    \tilde f_1 f_3 <0,
    \label{appendixFHN:cubic-condition}
\end{eqnarray}
is satisfied, then the cubic fast equation can be rescaled to the canonical FHN form. To do so, introduce the rescaled time
\begin{eqnarray}
    \theta=\tilde f_1 t \, \Rightarrow \frac{\mathrm{d}}{\mathrm{d} t} = \frac{\mathrm{d}\theta}{\mathrm{d} t} \frac{\mathrm{d}}{\mathrm{d} \theta} = \tilde f_1 \frac{\mathrm{d}}{\mathrm{d} \theta} \, \, \mathrm{or} \, \, \frac{\mathrm{d}}{\mathrm{d} \theta} = \frac{1}{\tilde f_1} \frac{\mathrm{d}}{\mathrm{d} t},
\end{eqnarray}
and choosing
\begin{eqnarray}\label{eq:lambda_square}
V=\lambda x, \qquad \lambda^2=-\frac{3f_3}{\tilde f_1},
\end{eqnarray}
we obtain
\begin{eqnarray}\nonumber
&&\frac{\mathrm{d}V}{\mathrm{d}\theta} = \frac{1}{\tilde{f}_1} \frac{\mathrm{d}}{\mathrm{d}t} (\lambda x) =  \frac{1}{\tilde{f}_1} \lambda \dot{x} = \\ \nonumber   
&& = \frac{\lambda}{\tilde{f}_1} (\tilde f_0+\tilde f_1 x+f_3 x^3+f_w\eta + \ldots)\\ \nonumber
&& = \frac{\lambda\tilde f_0}{\tilde{f}_1}  + V + \frac{f_3}{\tilde{f}_1} \left( \frac{1}{\lambda^2} \right) V^3 + \frac{\lambda f_w}{\tilde{f}_1} \eta + \ldots \, .\\ \nonumber
\end{eqnarray}

Using Eq.(\ref{eq:lambda_square}), we thus obtain our final result
\begin{eqnarray}\label{appendixFHN:FHN-fast-derived}
\frac{dV}{d\theta}=V-\frac{V^3}{3}-W+I_0+\mathcal O(V^4,VW,W^2),   
\end{eqnarray}
where
\begin{eqnarray}\label{params:W_I0}
W=-\frac{\lambda f_w}{\tilde f_1}\eta, \qquad I_0=\frac{\lambda \tilde f_0}{\tilde f_1}.
\end{eqnarray}
Thus, the FHN fast equation is a local normal form of the projected N\'eel dynamics. 
\\
\\
The recovery equation Eq.(\ref{appendixFHN:G-vw})   follows from the expansion of $G(v,w)$. To first order,
\begin{eqnarray}
    \dot \eta =
    g_0+g_v \xi+g_w\eta+\mathcal O(\xi^2,\xi\eta,\eta^2),
    \label{appendixFHN:recovery-linear}
\end{eqnarray}
where we introduced the notations:
\begin{eqnarray}
&&g_0 = G(v_\ast,w_\ast),\nonumber\\
&&g_v = G_v(v_\ast,w_\ast),\\
&&g_w = G_w(v_\ast,w_\ast).\nonumber
\end{eqnarray}
Here, we explicitly calculate $G_v(v_{\ast}, w_{\ast})$ and $G_w(v_{\ast}, w_{\ast})$

\begin{eqnarray} \nonumber
G_v(v_\ast,w_\ast)&&= \frac{\partial}{\partial v} \left\{ -\frac{1}{\alpha} \left[ C \, s \, \left(v \sqrt{2w-w^2-v^2} + \right) \right. \right. \\ \nonumber
&&  \left. \left. + \tau p_z (2w - w^2) \right] \right\} \\ \nonumber
&& = -\frac{1}{\alpha} \left\{ C s \left[ 1 \cdot \sqrt{2w-w^2-v^2} + \right. \right.  \\ \nonumber 
&& \left. \left. + \frac{(-2 v) v}{2\sqrt{2w-w^2-v^2}} \right] + \tau p_z \frac{\partial}{\partial v}(2 w - w^2) \right\}\\ \nonumber
&& = - \frac{1}{\alpha} \left\{ C s \left[ \frac{y}{s} - \frac{v^2}{(y/s)} \right] + 0 \right\}\\ \nonumber
&& = - \frac{1}{\alpha} \left[ C \, y - C \frac{v^2}{y} \right]\\ 
&&=- \frac{C}{\alpha} \left[ y - \frac{v^2}{y} \right] ,\\ \nonumber
G_w(v_\ast,w_\ast)&&= \frac{\partial}{\partial w}\left\{-\frac{1}{\alpha}\left[ C v\,s\sqrt{2w-w^2-v^2}+ \right. \right. \\ \nonumber 
&& \left. + \tau p_z(2w-w^2) \big] \right\} \\ \nonumber
&&= - \frac{1}{\alpha}\left\{ C v s \frac{1}{2} \frac{2 (1 - w)}{\sqrt{2w-w^2-v^2}} + \right. \\ \nonumber
&& \left.   + 2 (1 - w)\tau \, p_z\right\}\\ \nonumber
&&= - \frac{1}{\alpha}\left\{ C v \frac{(1 -w)}{y} + 2 (1 - w)\tau \, p_z\right\}\\ 
&&= -\frac{(1 - w)}{\alpha} \left[ C \frac{v}{y} + 2 \, \tau \, p_z \right].
\end{eqnarray}
Using again the shift \(\xi=x+h\),~Eq.(\ref{appendixFHN:recovery-linear}) becomes
\begin{eqnarray}\nonumber
\dot\eta  && = g_0+g_v (x + h) +g_w\eta + \cdots \\ 
&& = (g_0 + g_v \, h) + g_v x + g_w \eta + \cdots \\ \nonumber  
&&=\tilde g_0+g_v x+g_w\eta+\cdots .
\end{eqnarray}
where 
\begin{equation*}
\tilde g_0=g_0+g_v h  \, .   
\end{equation*}

Equivalently, following the same recipe as before, in order to put the recovery variable in terms of the normalized variables $V$, $W$, and $\theta$ and obtain
\begin{eqnarray}\label{appendixFHN:FHN-recovery-derived}
\frac{dW}{d\theta}=\epsilon(V+a-bW)+\cdots,   
\end{eqnarray}

we take into account 

\begin{equation}
\left.
\begin{aligned}
\frac{\mathrm{d}}{\mathrm{d} t} &= \frac{\mathrm{d}\theta}{\mathrm{d} t}\frac{\mathrm{d}}{\mathrm{d} \theta} = \tilde{f_1} \frac{\mathrm{d}}{\mathrm{d} \theta}\\
\eta &= - \frac{\tilde{f_1}}{\lambda f_w} \mathrm{W}\\
\end{aligned}\right\}  \Rightarrow  \dot{\eta} = - \frac{\tilde{f_1}^2}{\lambda f_w} \frac{\mathrm{d W}}{\mathrm{d} \theta},
\end{equation}

and given also that $ x = \frac{\mathrm{V}}{\lambda}$, the shifted equation takes the form

\begin{equation} \nonumber 
\begin{aligned}
& \dot\eta =\tilde g_0+g_v x+g_w\eta+\cdots \\ \nonumber
& \Rightarrow - \frac{\tilde{f_1}^2}{\lambda f_w} \frac{\mathrm{d W}}{\mathrm{d} \theta} = \tilde g_0 + \frac{g_v}{ \lambda} \mathrm{V} + g_w \eta + \ldots .
\end{aligned}
\end{equation}

\begin{eqnarray}
\frac{\mathrm{d W}}{\mathrm{d} \theta} &&= - \frac{\lambda f_w \tilde g_0}{\tilde{f_1}^2} - \frac{f_w  g_v}{\tilde{f_1}^2} \mathrm{V} + \frac{g_w}{\tilde{f_1}} \mathrm{W}\\ \nonumber
&& = - \frac{f_w  g_v}{\tilde{f_1}^2} \left[\mathrm{V} + \frac{\lambda \, \tilde{g_0}}{g_v}  - \frac{g_w}{\tilde{f_1}} \left(\frac{ \tilde{f_1}^2}{f_w g_v}\right) \mathrm{W}  \right]. \\ \nonumber
\end{eqnarray}

The obtained result can be written as

\begin{equation}
\frac{\mathrm{d W}}{\mathrm{d} \theta} = \epsilon \left( \mathrm{V} + a - b \, \mathrm{W} \right),
\end{equation}
where
\begin{eqnarray}
&&\epsilon \equiv - \frac{f_w  g_v}{\tilde{f_1}^2}, \\ \nonumber
&& a \equiv \frac{\lambda \, \tilde{g_0}}{g_v}, \\ \nonumber
&& b \equiv -\frac{g_w}{\epsilon \tilde{f_1}}. \\ \nonumber
\end{eqnarray}

\section{Fokker-Planck equation}

Starting from Eq.~(66), write the stochastic system in the compact form
\begin{align}
&dV_\theta = A_V(V_\theta,W_\theta,\theta)\,d\theta+\sigma_V\,dB_V(\theta),\\
&dW_\theta = A_W(V_\theta,W_\theta)\,d\theta +\sigma_W\,dB_W(\theta),
\end{align}
where
\begin{align}
&A_V(V,W,\theta)=V-\frac{V^3}{3}-W+I_0+ \chi_{\rm sp}I_{\rm sp}^{\rm ext}(\theta),\\
& A_W(V,W)=\epsilon(V+a-bW).
\end{align}
Let $P(V,W,\theta)$ be the probability density of the random vector $(V_\theta,W_\theta)$. Let also $\phi(V,W)$ be a smooth test function.
Using It\^o's formula, the backward generator acting on $\phi$ is
\begin{align}
&\mathcal L\phi=A_V\frac{\partial \phi}{\partial V}+A_W\frac{\partial \phi}{\partial W}+
\frac{\sigma_V^2}{2}
\frac{\partial^2\phi}{\partial V^2}+\frac{\sigma_W^2}{2}
\frac{\partial^2\phi}{\partial W^2}.
\end{align}
Taking expectation gives the weak form
\begin{align}
&\frac{d}{d\theta}\int_{\mathbb R^2}\phi(V,W)P(V,W,\theta)\,dV\,dW=\nonumber\\
&=\int_{\mathbb R^2}(\mathcal L\phi)(V,W)P(V,W,\theta)\,dV\,dW.
\end{align}
Since $\phi$ does not explicitly depend on $\theta$, the left-hand side is
\begin{eqnarray}
\frac{d}{d\theta}\int_{\mathbb R^2}\phi P\,dV\,dW=\int_{\mathbb R^2}\phi\frac{\partial P}{\partial\theta}\,dV\,dW.
\end{eqnarray}
Therefore,
\begin{align}
&\int_{\mathbb R^2}\phi\frac{\partial P}{\partial\theta}\,dV\,dW=
\int_{\mathbb R^2}P A_V\frac{\partial\phi}{\partial V}\,dV\,dW+
\nonumber\\
&+\int_{\mathbb R^2}P A_W\frac{\partial\phi}{\partial W}\,dV\,dW+\frac{\sigma_V^2}{2}\int_{\mathbb R^2}P\frac{\partial^2\phi}{\partial V^2}\,dV\,dW  \nonumber \\
&+\frac{\sigma_W^2}{2}\int_{\mathbb R^2}P\frac{\partial^2\phi}{\partial W^2}\,dV\,dW.
\end{align}
We now move the derivatives from the test function $\phi$ onto the density $P$. This is precisely the passage from the backward generator
$\mathcal L$ to its adjoint, or forward, generator $\mathcal L^*$. For clarity, we treat each term separately.

\subsubsection*{1. Drift term in the $V$-direction}

We consider the integral
\begin{eqnarray}
I_V=\int_{\mathbb R^2}P A_V \frac{\partial\phi}{\partial V}\,dV\,dW.
\end{eqnarray}
For fixed $W$, we integrate by parts with respect to $V$:
\begin{align}
&I_V =\int_{-\infty}^{\infty}\left[\int_{-\infty}^{\infty}A_V(V,W,\theta)P(V,W,\theta)\times \right. \nonumber \\
& \quad \, \, \times \left. \frac{\partial\phi}{\partial V}(V,W)\,dV\right]dW\\
&  =\int_{-\infty}^{\infty}\left[\left.A_VP\phi\right|_{V=-\infty}^{V=\infty}-
\int_{-\infty}^{\infty}\phi\frac{\partial}{\partial V}(A_VP)\,dV\right]dW.\nonumber
\end{align}
Assuming that the boundary term vanishes,
\begin{eqnarray}
\left.A_VP\phi\right|_{V=-\infty}^{V=\infty}=0,
\end{eqnarray}
we obtain
\begin{eqnarray}
I_V=-\int_{\mathbb R^2}\phi\frac{\partial}{\partial V}(A_VP)\,dV\,dW.
\end{eqnarray}
Thus,
\begin{align}
&\int_{\mathbb R^2}P A_V \frac{\partial\phi}{\partial V}\,dV\,dW= \nonumber \\
& =-\int_{\mathbb R^2}\phi\frac{\partial}{\partial V}(A_VP)\,dV\,dW.
\end{align}
This fact explains why the Fokker--Planck equation contains the  term
\begin{eqnarray}
-\frac{\partial}{\partial V}(A_VP),
\end{eqnarray}
and  not merely
\begin{eqnarray}
-A_V\frac{\partial P}{\partial V}.
\end{eqnarray}
Taking into account
\begin{eqnarray}
-\frac{\partial}{\partial V}(A_VP)=-\frac{\partial A_V}{\partial V}P-
A_V\frac{\partial P}{\partial V}.
\end{eqnarray}
For our expression of the drift,
\begin{eqnarray}
A_V=V-\frac{V^3}{3}-W+I_0+\chi_{\rm sp}I_{\rm sp}^{\rm ext}(\theta),
\end{eqnarray}
one obtains
\begin{eqnarray}
\frac{\partial A_V}{\partial V}=1-V^2.
\end{eqnarray}
Therefore,
\begin{eqnarray}
&&-\frac{\partial}{\partial V}(A_VP)=-(1-V^2)P-\\
&&-\left(V-\frac{V^3}{3}-W+I_0+\chi_{\rm sp}I_{\rm sp}^{\rm ext}(\theta)
\right)\frac{\partial P}{\partial V}.\nonumber
\end{eqnarray}
The compact divergence form is usually preferred:
\begin{eqnarray}
-\frac{\partial}{\partial V}\left[\left(V-\frac{V^3}{3}-W+I_0+\chi_{\rm sp}I_{\rm sp}^{\rm ext}(\theta)\right)P\right].
\end{eqnarray}

\subsubsection*{2. Drift term in the $W$-direction}

Now we proceed to the term
\begin{eqnarray}
I_W=\int_{\mathbb R^2}P A_W \frac{\partial\phi}{\partial W}\,dV\,dW.
\end{eqnarray}
For fixed $V$, integrate by parts with respect to $W$:
\begin{align}
&I_W =\int_{-\infty}^{\infty}\left[\int_{-\infty}^{\infty}A_W(V,W)P(V,W,\theta)\times \right.\nonumber \\
& \quad \, \,  \left.\times \frac{\partial\phi}{\partial W}(V,W)dW\right]dV \\
&=\int_{-\infty}^{\infty}\left[\left.A_WP\phi\right|_{W=-\infty}^{W=\infty}-\int_{-\infty}^{\infty}\phi\frac{\partial}{\partial W}(A_WP)dW\right]dV. \nonumber
\end{align}
Assuming the boundary term vanishes,
\begin{eqnarray}
\left.A_WP\phi\right|_{W=-\infty}^{W=\infty}=0,
\end{eqnarray}
we get
\begin{eqnarray}
I_W=-\int_{\mathbb R^2}\phi\frac{\partial}{\partial W}(A_WP)\,dV\,dW.
\end{eqnarray}
Therefore,
\begin{eqnarray}
&&\int_{\mathbb R^2}P A_W \frac{\partial\phi}{\partial W}\,dV\,dW=\\
&&=-\int_{\mathbb R^2}\phi\frac{\partial}{\partial W}(A_WP)\,dV\,dW.\nonumber
\end{eqnarray}
Since
\begin{eqnarray}
A_W=\epsilon(V+a-bW),
\end{eqnarray}
we have
\begin{eqnarray}
\frac{\partial A_W}{\partial W}=-\epsilon b.
\end{eqnarray}
Hence, if expanded,
\begin{eqnarray}
&&-\frac{\partial}{\partial W}(A_WP)=-\frac{\partial A_W}{\partial W}P-A_W\frac{\partial P}{\partial W}=\nonumber\\
&& = \epsilon b P-\epsilon(V+a-bW)\frac{\partial P}{\partial W}.
\end{eqnarray}
Again, the compact divergence form reads 
\begin{eqnarray}
-\frac{\partial}{\partial W}\left[\epsilon(V+a-bW)P\right].
\end{eqnarray}

\subsubsection*{3. Diffusion term in the $V$-direction}

Now we consider the second-order term
\begin{align}
&I_{VV}=\frac{\sigma_V^2}{2}\int_{\mathbb R^2}P\frac{\partial^2\phi}{\partial V^2}\,dV\,dW.
\end{align}
For fixed $W$, integrate by parts once with respect to $V$:
\begin{align}
&I_{VV}=\frac{\sigma_V^2}{2}\int_{-\infty}^{\infty}\left[\int_{-\infty}^{\infty}P\frac{\partial^2\phi}{\partial V^2}\,dV\right]dW=\\
&=\frac{\sigma_V^2}{2}\int_{-\infty}^{\infty}\left[\left.P\frac{\partial\phi}{\partial V}\right|_{V=-\infty}^{V=\infty}-\int_{-\infty}^{\infty}
\frac{\partial P}{\partial V}\frac{\partial\phi}{\partial V}\,dV\right]dW.\nonumber
\end{align}
We assume that
\begin{align}
\left.P\frac{\partial\phi}{\partial V}\right|_{V=-\infty}^{V=\infty}=0.
\end{align}
Then
\begin{eqnarray}
I_{VV}=-\frac{\sigma_V^2}{2}\int_{\mathbb R^2}\frac{\partial P}{\partial V}\frac{\partial\phi}{\partial V}\,dV\,dW.
\end{eqnarray}
Integrate by parts a second time:
\begin{align}
&I_{VV}=-\frac{\sigma_V^2}{2}\int_{-\infty}^{\infty}\left[\int_{-\infty}^{\infty}\frac{\partial P}{\partial V}\frac{\partial\phi}{\partial V}\,dV\right]dW=\\
&=-\frac{\sigma_V^2}{2}\int_{-\infty}^{\infty}\left[\left.\frac{\partial P}{\partial V}\phi\right|_{V=-\infty}^{V=\infty}-
\int_{-\infty}^{\infty}\phi\frac{\partial^2P}{\partial V^2}\,dV\right]dW.\nonumber
\end{align}
Let us assume
\begin{eqnarray}
\left.\frac{\partial P}{\partial V}\phi\right|_{V=-\infty}^{V=\infty}=0.
\end{eqnarray}
Therefore,
\begin{eqnarray}
I_{VV}=\frac{\sigma_V^2}{2}\int_{\mathbb R^2}\phi\frac{\partial^2P}{\partial V^2}\,dV\,dW.
\end{eqnarray}
Thus,
\begin{align}
&\frac{\sigma_V^2}{2}\int_{\mathbb R^2}P\frac{\partial^2\phi}{\partial V^2}\,dV\,dW=\\
&=\frac{\sigma_V^2}{2}\int_{\mathbb R^2}\phi\frac{\partial^2P}{\partial V^2}\,dV\,dW.\nonumber
\end{align}
This is why the second derivative keeps the same sign in the Fokker--Planck equation:
\begin{eqnarray}
+\frac{\sigma_V^2}{2}\frac{\partial^2P}{\partial V^2}.
\end{eqnarray}

\subsubsection*{4. Diffusion term in the $W$-direction}

Similarly,to previous subsection 
\begin{eqnarray}
I_{WW}=\frac{\sigma_W^2}{2}\int_{\mathbb R^2}P\frac{\partial^2\phi}{\partial W^2}\,dV\,dW.
\end{eqnarray}
Integrating by parts twice with respect to $W$, and assuming that the boundary terms vanish, gives
\begin{align}
&\frac{\sigma_W^2}{2}\int_{\mathbb R^2}P\frac{\partial^2\phi}{\partial W^2}\,dV\,dW=\\
&=\frac{\sigma_W^2}{2}\int_{\mathbb R^2}\phi\frac{\partial^2P}{\partial W^2}\,dV\,dW.
\end{align}
Therefore, the Fokker--Planck equation contains
\begin{equation}
+\frac{\sigma_W^2}{2}\frac{\partial^2P}{\partial W^2}.
\end{equation}

\subsubsection*{5. Combining all terms}

Putting the four transformed pieces together leads to
\begin{align}
&\int_{\mathbb R^2}\phi\frac{\partial P}{\partial\theta}\,dV\,dW=-\int_{\mathbb R^2}\phi\frac{\partial}{\partial V}(A_VP)\,dV\,dW-\nonumber\\
&-\int_{\mathbb R^2}\phi\frac{\partial}{\partial W}(A_WP)\,dV\,dW+\frac{\sigma_V^2}{2}\int_{\mathbb R^2}\phi\frac{\partial^2P}{\partial V^2}\,dV\,dW\nonumber\\
&+\frac{\sigma_W^2}{2}\int_{\mathbb R^2}\phi\frac{\partial^2P}{\partial W^2}\,dV\,dW.
\end{align}
Equivalently,
\begin{align}
&\int_{\mathbb R^2}\phi\big[\frac{\partial P}{\partial\theta}+\frac{\partial}{\partial V}(A_VP)+\frac{\partial}{\partial W}(A_WP)-\\
&-\frac{\sigma_V^2}{2}\frac{\partial^2P}{\partial V^2}-\frac{\sigma_W^2}{2}\frac{\partial^2P}{\partial W^2}\big]dV\,dW=0.\nonumber
\end{align}
Since this identity holds for every smooth test function $\phi$, the quantity inside the square brackets must vanish. Therefore,
\begin{eqnarray}
&&\frac{\partial P}{\partial\theta}+\frac{\partial}{\partial V}(A_VP)+\frac{\partial}{\partial W}(A_WP)
-\frac{\sigma_V^2}{2}\frac{\partial^2P}{\partial V^2}\nonumber\\
&&-\frac{\sigma_W^2}{2}\frac{\partial^2P}{\partial W^2}=0.
\end{eqnarray}
Equivalently,
\begin{align}
\frac{\partial P}{\partial\theta}&=-\frac{\partial}{\partial V}(A_VP)-\frac{\partial}{\partial W}(A_WP)+\frac{\sigma_V^2}{2}
\frac{\partial^2P}{\partial V^2}+\nonumber\\
&+\frac{\sigma_W^2}{2}\frac{\partial^2P}{\partial W^2}.
\end{align}
Substituting the explicit drifts,
\begin{align}
&A_V(V,W,\theta)=V-\frac{V^3}{3}-W+I_0+\chi_{\rm sp}I_{\rm sp}^{\rm ext}(\theta),\nonumber\\
&A_W(V,W)=\epsilon(V+a-bW),
\end{align}
we finally obtain
\begin{align}
&\frac{\partial P}{\partial \theta}=-\frac{\partial}{\partial V}\left\{\left[V-\frac{V^3}{3}-W
+I_0+\chi_{\rm sp}I_{\rm sp}^{\rm ext}(\theta)\right]P\right\}-\nonumber\\
&\frac{\partial}{\partial W}\left\{\epsilon(V+a-bW)P\right\}+\frac{\sigma_V^2}{2}\frac{\partial^2P}{\partial V^2}+\frac{\sigma_W^2}{2}\frac{\partial^2P}{\partial W^2}.
\end{align}

\bibliography{Dimitris1}

\begin{thebibliography}{33}%
\makeatletter
\providecommand \@ifxundefined [1]{%
 \@ifx{#1\undefined}
}%
\providecommand \@ifnum [1]{%
 \ifnum #1\expandafter \@firstoftwo
 \else \expandafter \@secondoftwo
 \fi
}%
\providecommand \@ifx [1]{%
 \ifx #1\expandafter \@firstoftwo
 \else \expandafter \@secondoftwo
 \fi
}%
\providecommand \natexlab [1]{#1}%
\providecommand \enquote  [1]{``#1''}%
\providecommand \bibnamefont  [1]{#1}%
\providecommand \bibfnamefont [1]{#1}%
\providecommand \citenamefont [1]{#1}%
\providecommand \href@noop [0]{\@secondoftwo}%
\providecommand \href [0]{\begingroup \@sanitize@url \@href}%
\providecommand \@href[1]{\@@startlink{#1}\@@href}%
\providecommand \@@href[1]{\endgroup#1\@@endlink}%
\providecommand \@sanitize@url [0]{\catcode `\\12\catcode `\$12\catcode
  `\&12\catcode `\#12\catcode `\^12\catcode `\_12\catcode `\%12\relax}%
\providecommand \@@startlink[1]{}%
\providecommand \@@endlink[0]{}%
\providecommand \url  [0]{\begingroup\@sanitize@url \@url }%
\providecommand \@url [1]{\endgroup\@href {#1}{\urlprefix }}%
\providecommand \urlprefix  [0]{URL }%
\providecommand \Eprint [0]{\href }%
\providecommand \doibase [0]{https://doi.org/}%
\providecommand \selectlanguage [0]{\@gobble}%
\providecommand \bibinfo  [0]{\@secondoftwo}%
\providecommand \bibfield  [0]{\@secondoftwo}%
\providecommand \translation [1]{[#1]}%
\providecommand \BibitemOpen [0]{}%
\providecommand \bibitemStop [0]{}%
\providecommand \bibitemNoStop [0]{.\EOS\space}%
\providecommand \EOS [0]{\spacefactor3000\relax}%
\providecommand \BibitemShut  [1]{\csname bibitem#1\endcsname}%
\let\auto@bib@innerbib\@empty
\bibitem [{\citenamefont {Zhu}\ \emph {et~al.}(2020)\citenamefont {Zhu},
  \citenamefont {Zhang}, \citenamefont {Yang},\ and\ \citenamefont
  {Huang}}]{zhu2020comprehensive}%
  \BibitemOpen
  \bibfield  {author} {\bibinfo {author} {\bibfnamefont {J.}~\bibnamefont
  {Zhu}}, \bibinfo {author} {\bibfnamefont {T.}~\bibnamefont {Zhang}}, \bibinfo
  {author} {\bibfnamefont {Y.}~\bibnamefont {Yang}},\ and\ \bibinfo {author}
  {\bibfnamefont {R.}~\bibnamefont {Huang}},\ }\bibfield  {title} {\bibinfo
  {title} {A comprehensive review on emerging artificial neuromorphic
  devices},\ }\href@noop {} {\bibfield  {journal} {\bibinfo  {journal} {Applied
  Physics Reviews}\ }\textbf {\bibinfo {volume} {7}} (\bibinfo {year}
  {2020})}\BibitemShut {NoStop}%
\bibitem [{\citenamefont {Wang}\ \emph {et~al.}(2017)\citenamefont {Wang},
  \citenamefont {Wang}, \citenamefont {Nagai}, \citenamefont {Xie},
  \citenamefont {Yi},\ and\ \citenamefont {Huang}}]{wang2017nanoionics}%
  \BibitemOpen
  \bibfield  {author} {\bibinfo {author} {\bibfnamefont {Z.}~\bibnamefont
  {Wang}}, \bibinfo {author} {\bibfnamefont {L.}~\bibnamefont {Wang}}, \bibinfo
  {author} {\bibfnamefont {M.}~\bibnamefont {Nagai}}, \bibinfo {author}
  {\bibfnamefont {L.}~\bibnamefont {Xie}}, \bibinfo {author} {\bibfnamefont
  {M.}~\bibnamefont {Yi}},\ and\ \bibinfo {author} {\bibfnamefont
  {W.}~\bibnamefont {Huang}},\ }\bibfield  {title} {\bibinfo {title}
  {Nanoionics-enabled memristive devices: strategies and materials for
  neuromorphic applications},\ }\href@noop {} {\bibfield  {journal} {\bibinfo
  {journal} {Advanced Electronic Materials}\ }\textbf {\bibinfo {volume} {3}},\
  \bibinfo {pages} {1600510} (\bibinfo {year} {2017})}\BibitemShut {NoStop}%
\bibitem [{\citenamefont {Gupta}\ \emph {et~al.}(2025)\citenamefont {Gupta},
  \citenamefont {Patel}, \citenamefont {Kumar}, \citenamefont {Pohl},
  \citenamefont {Park}, \citenamefont {Youn}, \citenamefont {Jeong},\ and\
  \citenamefont {Kim}}]{gupta2025toward}%
  \BibitemOpen
  \bibfield  {author} {\bibinfo {author} {\bibfnamefont {S.~U.}\ \bibnamefont
  {Gupta}}, \bibinfo {author} {\bibfnamefont {M.}~\bibnamefont {Patel}},
  \bibinfo {author} {\bibfnamefont {N.}~\bibnamefont {Kumar}}, \bibinfo
  {author} {\bibfnamefont {L.}~\bibnamefont {Pohl}}, \bibinfo {author}
  {\bibfnamefont {M.-J.}\ \bibnamefont {Park}}, \bibinfo {author}
  {\bibfnamefont {S.-M.}\ \bibnamefont {Youn}}, \bibinfo {author}
  {\bibfnamefont {C.}~\bibnamefont {Jeong}},\ and\ \bibinfo {author}
  {\bibfnamefont {J.}~\bibnamefont {Kim}},\ }\bibfield  {title} {\bibinfo
  {title} {Toward advancement of fabrication techniques of neuromorphic
  computing devices based on 2d materials},\ }\href@noop {} {\bibfield
  {journal} {\bibinfo  {journal} {Advanced Materials Technologies}\ }\textbf
  {\bibinfo {volume} {10}},\ \bibinfo {pages} {e00786} (\bibinfo {year}
  {2025})}\BibitemShut {NoStop}%
\bibitem [{\citenamefont {Aimone}\ \emph {et~al.}(2022)\citenamefont {Aimone},
  \citenamefont {Date}, \citenamefont {Fonseca-Guerra}, \citenamefont
  {Hamilton}, \citenamefont {Henke}, \citenamefont {Kay}, \citenamefont
  {Kenyon}, \citenamefont {Kulkarni}, \citenamefont {Mniszewski}, \citenamefont
  {Parsa} \emph {et~al.}}]{aimone2022review}%
  \BibitemOpen
  \bibfield  {author} {\bibinfo {author} {\bibfnamefont {J.~B.}\ \bibnamefont
  {Aimone}}, \bibinfo {author} {\bibfnamefont {P.}~\bibnamefont {Date}},
  \bibinfo {author} {\bibfnamefont {G.~A.}\ \bibnamefont {Fonseca-Guerra}},
  \bibinfo {author} {\bibfnamefont {K.~E.}\ \bibnamefont {Hamilton}}, \bibinfo
  {author} {\bibfnamefont {K.}~\bibnamefont {Henke}}, \bibinfo {author}
  {\bibfnamefont {B.}~\bibnamefont {Kay}}, \bibinfo {author} {\bibfnamefont
  {G.~T.}\ \bibnamefont {Kenyon}}, \bibinfo {author} {\bibfnamefont {S.~R.}\
  \bibnamefont {Kulkarni}}, \bibinfo {author} {\bibfnamefont {S.~M.}\
  \bibnamefont {Mniszewski}}, \bibinfo {author} {\bibfnamefont
  {M.}~\bibnamefont {Parsa}}, \emph {et~al.},\ }\bibfield  {title} {\bibinfo
  {title} {A review of non-cognitive applications for neuromorphic computing},\
  }\href@noop {} {\bibfield  {journal} {\bibinfo  {journal} {Neuromorphic
  Computing and Engineering}\ }\textbf {\bibinfo {volume} {2}},\ \bibinfo
  {pages} {032003} (\bibinfo {year} {2022})}\BibitemShut {NoStop}%
\bibitem [{\citenamefont {Mead}(2002)}]{mead2002neuromorphic}%
  \BibitemOpen
  \bibfield  {author} {\bibinfo {author} {\bibfnamefont {C.}~\bibnamefont
  {Mead}},\ }\bibfield  {title} {\bibinfo {title} {Neuromorphic electronic
  systems},\ }\href@noop {} {\bibfield  {journal} {\bibinfo  {journal}
  {Proceedings of the IEEE}\ }\textbf {\bibinfo {volume} {78}},\ \bibinfo
  {pages} {1629} (\bibinfo {year} {2002})}\BibitemShut {NoStop}%
\bibitem [{\citenamefont {Grollier}\ \emph {et~al.}(2016)\citenamefont
  {Grollier}, \citenamefont {Querlioz},\ and\ \citenamefont
  {Stiles}}]{grollier2016spintronic}%
  \BibitemOpen
  \bibfield  {author} {\bibinfo {author} {\bibfnamefont {J.}~\bibnamefont
  {Grollier}}, \bibinfo {author} {\bibfnamefont {D.}~\bibnamefont {Querlioz}},\
  and\ \bibinfo {author} {\bibfnamefont {M.~D.}\ \bibnamefont {Stiles}},\
  }\bibfield  {title} {\bibinfo {title} {Spintronic nanodevices for bioinspired
  computing},\ }\href@noop {} {\bibfield  {journal} {\bibinfo  {journal}
  {Proceedings of the IEEE}\ }\textbf {\bibinfo {volume} {104}},\ \bibinfo
  {pages} {2024} (\bibinfo {year} {2016})}\BibitemShut {NoStop}%
\bibitem [{\citenamefont {Torrejon}\ \emph {et~al.}(2017)\citenamefont
  {Torrejon}, \citenamefont {Riou}, \citenamefont {Araujo}, \citenamefont
  {Tsunegi}, \citenamefont {Khalsa}, \citenamefont {Querlioz}, \citenamefont
  {Bortolotti}, \citenamefont {Cros}, \citenamefont {Yakushiji}, \citenamefont
  {Fukushima} \emph {et~al.}}]{torrejon2017neuromorphic}%
  \BibitemOpen
  \bibfield  {author} {\bibinfo {author} {\bibfnamefont {J.}~\bibnamefont
  {Torrejon}}, \bibinfo {author} {\bibfnamefont {M.}~\bibnamefont {Riou}},
  \bibinfo {author} {\bibfnamefont {F.~A.}\ \bibnamefont {Araujo}}, \bibinfo
  {author} {\bibfnamefont {S.}~\bibnamefont {Tsunegi}}, \bibinfo {author}
  {\bibfnamefont {G.}~\bibnamefont {Khalsa}}, \bibinfo {author} {\bibfnamefont
  {D.}~\bibnamefont {Querlioz}}, \bibinfo {author} {\bibfnamefont
  {P.}~\bibnamefont {Bortolotti}}, \bibinfo {author} {\bibfnamefont
  {V.}~\bibnamefont {Cros}}, \bibinfo {author} {\bibfnamefont {K.}~\bibnamefont
  {Yakushiji}}, \bibinfo {author} {\bibfnamefont {A.}~\bibnamefont
  {Fukushima}}, \emph {et~al.},\ }\bibfield  {title} {\bibinfo {title}
  {Neuromorphic computing with nanoscale spintronic oscillators},\ }\href@noop
  {} {\bibfield  {journal} {\bibinfo  {journal} {Nature}\ }\textbf {\bibinfo
  {volume} {547}},\ \bibinfo {pages} {428} (\bibinfo {year}
  {2017})}\BibitemShut {NoStop}%
\bibitem [{\citenamefont {Godinho}\ \emph {et~al.}(2024)\citenamefont
  {Godinho}, \citenamefont {Rout}, \citenamefont {Salikhov}, \citenamefont
  {Hellwig}, \citenamefont {{\v{S}}ob{\'a}{\v{n}}}, \citenamefont {Otxoa},
  \citenamefont {Olejn{\'\i}k}, \citenamefont {Jungwirth},\ and\ \citenamefont
  {Wunderlich}}]{godinho2024antiferromagnetic}%
  \BibitemOpen
  \bibfield  {author} {\bibinfo {author} {\bibfnamefont {J.}~\bibnamefont
  {Godinho}}, \bibinfo {author} {\bibfnamefont {P.}~\bibnamefont {Rout}},
  \bibinfo {author} {\bibfnamefont {R.}~\bibnamefont {Salikhov}}, \bibinfo
  {author} {\bibfnamefont {O.}~\bibnamefont {Hellwig}}, \bibinfo {author}
  {\bibfnamefont {Z.}~\bibnamefont {{\v{S}}ob{\'a}{\v{n}}}}, \bibinfo {author}
  {\bibfnamefont {R.}~\bibnamefont {Otxoa}}, \bibinfo {author} {\bibfnamefont
  {K.}~\bibnamefont {Olejn{\'\i}k}}, \bibinfo {author} {\bibfnamefont
  {T.}~\bibnamefont {Jungwirth}},\ and\ \bibinfo {author} {\bibfnamefont
  {J.}~\bibnamefont {Wunderlich}},\ }\bibfield  {title} {\bibinfo {title}
  {Antiferromagnetic domain wall memory with neuromorphic functionality},\
  }\href@noop {} {\bibfield  {journal} {\bibinfo  {journal} {npj Spintronics}\
  }\textbf {\bibinfo {volume} {2}},\ \bibinfo {pages} {39} (\bibinfo {year}
  {2024})}\BibitemShut {NoStop}%
\bibitem [{\citenamefont {Han}\ \emph {et~al.}(2024)\citenamefont {Han},
  \citenamefont {Wang}, \citenamefont {Wang}, \citenamefont {Wang},
  \citenamefont {Zheng}, \citenamefont {Zhao}, \citenamefont {Huang},
  \citenamefont {Cao}, \citenamefont {Chen}, \citenamefont {Bai} \emph
  {et~al.}}]{han2024neuromorphic}%
  \BibitemOpen
  \bibfield  {author} {\bibinfo {author} {\bibfnamefont {X.}~\bibnamefont
  {Han}}, \bibinfo {author} {\bibfnamefont {Z.}~\bibnamefont {Wang}}, \bibinfo
  {author} {\bibfnamefont {Y.}~\bibnamefont {Wang}}, \bibinfo {author}
  {\bibfnamefont {D.}~\bibnamefont {Wang}}, \bibinfo {author} {\bibfnamefont
  {L.}~\bibnamefont {Zheng}}, \bibinfo {author} {\bibfnamefont
  {L.}~\bibnamefont {Zhao}}, \bibinfo {author} {\bibfnamefont {Q.}~\bibnamefont
  {Huang}}, \bibinfo {author} {\bibfnamefont {Q.}~\bibnamefont {Cao}}, \bibinfo
  {author} {\bibfnamefont {Y.}~\bibnamefont {Chen}}, \bibinfo {author}
  {\bibfnamefont {L.}~\bibnamefont {Bai}}, \emph {et~al.},\ }\bibfield  {title}
  {\bibinfo {title} {Neuromorphic computing in synthetic antiferromagnets by
  spin-orbit torque induced magnetic-field-free magnetization switching},\
  }\href@noop {} {\bibfield  {journal} {\bibinfo  {journal} {Advanced
  Functional Materials}\ }\textbf {\bibinfo {volume} {34}},\ \bibinfo {pages}
  {2404679} (\bibinfo {year} {2024})}\BibitemShut {NoStop}%
\bibitem [{\citenamefont {Zhou}\ and\ \citenamefont
  {Chen}(2021)}]{zhou2021prospect}%
  \BibitemOpen
  \bibfield  {author} {\bibinfo {author} {\bibfnamefont {J.}~\bibnamefont
  {Zhou}}\ and\ \bibinfo {author} {\bibfnamefont {J.}~\bibnamefont {Chen}},\
  }\bibfield  {title} {\bibinfo {title} {Prospect of spintronics in
  neuromorphic computing},\ }\href@noop {} {\bibfield  {journal} {\bibinfo
  {journal} {Advanced Electronic Materials}\ }\textbf {\bibinfo {volume} {7}},\
  \bibinfo {pages} {2100465} (\bibinfo {year} {2021})}\BibitemShut {NoStop}%
\bibitem [{\citenamefont {Fukami}\ and\ \citenamefont
  {Ohno}(2018)}]{fukami2018perspective}%
  \BibitemOpen
  \bibfield  {author} {\bibinfo {author} {\bibfnamefont {S.}~\bibnamefont
  {Fukami}}\ and\ \bibinfo {author} {\bibfnamefont {H.}~\bibnamefont {Ohno}},\
  }\bibfield  {title} {\bibinfo {title} {Perspective: Spintronic synapse for
  artificial neural network},\ }\href@noop {} {\bibfield  {journal} {\bibinfo
  {journal} {Journal of Applied Physics}\ }\textbf {\bibinfo {volume} {124}}
  (\bibinfo {year} {2018})}\BibitemShut {NoStop}%
\bibitem [{\citenamefont {Sulymenko}\ \emph {et~al.}(2018)\citenamefont
  {Sulymenko}, \citenamefont {Prokopenko}, \citenamefont {Lisenkov},
  \citenamefont {{\AA}kerman}, \citenamefont {Tyberkevych}, \citenamefont
  {Slavin},\ and\ \citenamefont {Khymyn}}]{sulymenko2018ultra}%
  \BibitemOpen
  \bibfield  {author} {\bibinfo {author} {\bibfnamefont {O.}~\bibnamefont
  {Sulymenko}}, \bibinfo {author} {\bibfnamefont {O.}~\bibnamefont
  {Prokopenko}}, \bibinfo {author} {\bibfnamefont {I.}~\bibnamefont
  {Lisenkov}}, \bibinfo {author} {\bibfnamefont {J.}~\bibnamefont
  {{\AA}kerman}}, \bibinfo {author} {\bibfnamefont {V.}~\bibnamefont
  {Tyberkevych}}, \bibinfo {author} {\bibfnamefont {A.~N.}\ \bibnamefont
  {Slavin}},\ and\ \bibinfo {author} {\bibfnamefont {R.}~\bibnamefont
  {Khymyn}},\ }\bibfield  {title} {\bibinfo {title} {Ultra-fast logic devices
  using artificial “neurons” based on antiferromagnetic pulse generators},\
  }\href@noop {} {\bibfield  {journal} {\bibinfo  {journal} {Journal of Applied
  Physics}\ }\textbf {\bibinfo {volume} {124}} (\bibinfo {year}
  {2018})}\BibitemShut {NoStop}%
\bibitem [{\citenamefont {Ojha}\ \emph {et~al.}(2024)\citenamefont {Ojha},
  \citenamefont {Huang}, \citenamefont {Lin}, \citenamefont {Chatterjee},
  \citenamefont {Chang},\ and\ \citenamefont {Tseng}}]{ojha2024neuromorphic}%
  \BibitemOpen
  \bibfield  {author} {\bibinfo {author} {\bibfnamefont {D.~K.}\ \bibnamefont
  {Ojha}}, \bibinfo {author} {\bibfnamefont {Y.-H.}\ \bibnamefont {Huang}},
  \bibinfo {author} {\bibfnamefont {Y.-L.}\ \bibnamefont {Lin}}, \bibinfo
  {author} {\bibfnamefont {R.}~\bibnamefont {Chatterjee}}, \bibinfo {author}
  {\bibfnamefont {W.-Y.}\ \bibnamefont {Chang}},\ and\ \bibinfo {author}
  {\bibfnamefont {Y.-C.}\ \bibnamefont {Tseng}},\ }\bibfield  {title} {\bibinfo
  {title} {Neuromorphic computing with emerging antiferromagnetic ordering in
  spin--orbit torque devices},\ }\href@noop {} {\bibfield  {journal} {\bibinfo
  {journal} {Nano letters}\ }\textbf {\bibinfo {volume} {24}},\ \bibinfo
  {pages} {7706} (\bibinfo {year} {2024})}\BibitemShut {NoStop}%
\bibitem [{\citenamefont {Fukami}\ \emph {et~al.}(2020)\citenamefont {Fukami},
  \citenamefont {Lorenz},\ and\ \citenamefont
  {Gomonay}}]{fukami2020antiferromagnetic}%
  \BibitemOpen
  \bibfield  {author} {\bibinfo {author} {\bibfnamefont {S.}~\bibnamefont
  {Fukami}}, \bibinfo {author} {\bibfnamefont {V.~O.}\ \bibnamefont {Lorenz}},\
  and\ \bibinfo {author} {\bibfnamefont {O.}~\bibnamefont {Gomonay}},\
  }\bibfield  {title} {\bibinfo {title} {Antiferromagnetic spintronics},\
  }\href@noop {} {\bibfield  {journal} {\bibinfo  {journal} {Journal of Applied
  Physics}\ }\textbf {\bibinfo {volume} {128}} (\bibinfo {year}
  {2020})}\BibitemShut {NoStop}%
\bibitem [{\citenamefont {Kurenkov}\ \emph {et~al.}(2019)\citenamefont
  {Kurenkov}, \citenamefont {DuttaGupta}, \citenamefont {Zhang}, \citenamefont
  {Fukami}, \citenamefont {Horio},\ and\ \citenamefont
  {Ohno}}]{kurenkov2019artificial}%
  \BibitemOpen
  \bibfield  {author} {\bibinfo {author} {\bibfnamefont {A.}~\bibnamefont
  {Kurenkov}}, \bibinfo {author} {\bibfnamefont {S.}~\bibnamefont
  {DuttaGupta}}, \bibinfo {author} {\bibfnamefont {C.}~\bibnamefont {Zhang}},
  \bibinfo {author} {\bibfnamefont {S.}~\bibnamefont {Fukami}}, \bibinfo
  {author} {\bibfnamefont {Y.}~\bibnamefont {Horio}},\ and\ \bibinfo {author}
  {\bibfnamefont {H.}~\bibnamefont {Ohno}},\ }\bibfield  {title} {\bibinfo
  {title} {Artificial neuron and synapse realized in an
  antiferromagnet/ferromagnet heterostructure using dynamics of spin--orbit
  torque switching},\ }\href@noop {} {\bibfield  {journal} {\bibinfo  {journal}
  {Advanced Materials}\ }\textbf {\bibinfo {volume} {31}},\ \bibinfo {pages}
  {1900636} (\bibinfo {year} {2019})}\BibitemShut {NoStop}%
\bibitem [{\citenamefont {De~la Barrera}\ and\ \citenamefont
  {Nunez}(2025)}]{d6z2-t1sm}%
  \BibitemOpen
  \bibfield  {author} {\bibinfo {author} {\bibfnamefont {G.}~\bibnamefont
  {De~la Barrera}}\ and\ \bibinfo {author} {\bibfnamefont {A.~S.}\ \bibnamefont
  {Nunez}},\ }\bibfield  {title} {\bibinfo {title} {Nonlinear antiferromagnetic
  hall memristors},\ }\href {https://doi.org/10.1103/d6z2-t1sm} {\bibfield
  {journal} {\bibinfo  {journal} {Phys. Rev. B}\ }\textbf {\bibinfo {volume}
  {112}},\ \bibinfo {pages} {214452} (\bibinfo {year} {2025})}\BibitemShut
  {NoStop}%
\bibitem [{\citenamefont {Rodrigues}\ \emph {et~al.}(2023)\citenamefont
  {Rodrigues}, \citenamefont {Moukhader}, \citenamefont {Luo}, \citenamefont
  {Fang}, \citenamefont {Pontlevy}, \citenamefont {Hamadeh}, \citenamefont
  {Zeng}, \citenamefont {Carpentieri},\ and\ \citenamefont
  {Finocchio}}]{PhysRevApplied.19.064010}%
  \BibitemOpen
  \bibfield  {author} {\bibinfo {author} {\bibfnamefont {D.~R.}\ \bibnamefont
  {Rodrigues}}, \bibinfo {author} {\bibfnamefont {R.}~\bibnamefont
  {Moukhader}}, \bibinfo {author} {\bibfnamefont {Y.}~\bibnamefont {Luo}},
  \bibinfo {author} {\bibfnamefont {B.}~\bibnamefont {Fang}}, \bibinfo {author}
  {\bibfnamefont {A.}~\bibnamefont {Pontlevy}}, \bibinfo {author}
  {\bibfnamefont {A.}~\bibnamefont {Hamadeh}}, \bibinfo {author} {\bibfnamefont
  {Z.}~\bibnamefont {Zeng}}, \bibinfo {author} {\bibfnamefont {M.}~\bibnamefont
  {Carpentieri}},\ and\ \bibinfo {author} {\bibfnamefont {G.}~\bibnamefont
  {Finocchio}},\ }\bibfield  {title} {\bibinfo {title} {Spintronic
  hodgkin-huxley-analogue neuron implemented with a single magnetic tunnel
  junction},\ }\href {https://doi.org/10.1103/PhysRevApplied.19.064010}
  {\bibfield  {journal} {\bibinfo  {journal} {Phys. Rev. Appl.}\ }\textbf
  {\bibinfo {volume} {19}},\ \bibinfo {pages} {064010} (\bibinfo {year}
  {2023})}\BibitemShut {NoStop}%
\bibitem [{\citenamefont {Manna}\ \emph {et~al.}(2023)\citenamefont {Manna},
  \citenamefont {Medwal},\ and\ \citenamefont {Rawat}}]{PhysRevB.108.184411}%
  \BibitemOpen
  \bibfield  {author} {\bibinfo {author} {\bibfnamefont {S.}~\bibnamefont
  {Manna}}, \bibinfo {author} {\bibfnamefont {R.}~\bibnamefont {Medwal}},\ and\
  \bibinfo {author} {\bibfnamefont {R.~S.}\ \bibnamefont {Rawat}},\ }\bibfield
  {title} {\bibinfo {title} {Reconfigurable neural spiking in bias field free
  spin hall nano-oscillator},\ }\href
  {https://doi.org/10.1103/PhysRevB.108.184411} {\bibfield  {journal} {\bibinfo
   {journal} {Phys. Rev. B}\ }\textbf {\bibinfo {volume} {108}},\ \bibinfo
  {pages} {184411} (\bibinfo {year} {2023})}\BibitemShut {NoStop}%
\bibitem [{\citenamefont {Wang}\ \emph {et~al.}(2024)\citenamefont {Wang},
  \citenamefont {Csaba}, \citenamefont {Verba}, \citenamefont {Chumak},\ and\
  \citenamefont {Pirro}}]{PhysRevApplied.21.040503}%
  \BibitemOpen
  \bibfield  {author} {\bibinfo {author} {\bibfnamefont {Q.}~\bibnamefont
  {Wang}}, \bibinfo {author} {\bibfnamefont {G.}~\bibnamefont {Csaba}},
  \bibinfo {author} {\bibfnamefont {R.}~\bibnamefont {Verba}}, \bibinfo
  {author} {\bibfnamefont {A.~V.}\ \bibnamefont {Chumak}},\ and\ \bibinfo
  {author} {\bibfnamefont {P.}~\bibnamefont {Pirro}},\ }\bibfield  {title}
  {\bibinfo {title} {Nanoscale magnonic networks},\ }\href
  {https://doi.org/10.1103/PhysRevApplied.21.040503} {\bibfield  {journal}
  {\bibinfo  {journal} {Phys. Rev. Appl.}\ }\textbf {\bibinfo {volume} {21}},\
  \bibinfo {pages} {040503} (\bibinfo {year} {2024})}\BibitemShut {NoStop}%
\bibitem [{\citenamefont {Jungwirth}\ \emph {et~al.}(2016)\citenamefont
  {Jungwirth}, \citenamefont {Marti}, \citenamefont {Wadley},\ and\
  \citenamefont {Wunderlich}}]{jungwirth2016antiferromagnetic}%
  \BibitemOpen
  \bibfield  {author} {\bibinfo {author} {\bibfnamefont {T.}~\bibnamefont
  {Jungwirth}}, \bibinfo {author} {\bibfnamefont {X.}~\bibnamefont {Marti}},
  \bibinfo {author} {\bibfnamefont {P.}~\bibnamefont {Wadley}},\ and\ \bibinfo
  {author} {\bibfnamefont {J.}~\bibnamefont {Wunderlich}},\ }\bibfield  {title}
  {\bibinfo {title} {Antiferromagnetic spintronics},\ }\href@noop {} {\bibfield
   {journal} {\bibinfo  {journal} {Nature nanotechnology}\ }\textbf {\bibinfo
  {volume} {11}},\ \bibinfo {pages} {231} (\bibinfo {year} {2016})}\BibitemShut
  {NoStop}%
\bibitem [{\citenamefont {Gomonay}\ and\ \citenamefont
  {Loktev}(2014)}]{gomonay2014spintronics}%
  \BibitemOpen
  \bibfield  {author} {\bibinfo {author} {\bibfnamefont {E.}~\bibnamefont
  {Gomonay}}\ and\ \bibinfo {author} {\bibfnamefont {V.}~\bibnamefont
  {Loktev}},\ }\bibfield  {title} {\bibinfo {title} {Spintronics of
  antiferromagnetic systems},\ }\href@noop {} {\bibfield  {journal} {\bibinfo
  {journal} {Low Temperature Physics}\ }\textbf {\bibinfo {volume} {40}},\
  \bibinfo {pages} {17} (\bibinfo {year} {2014})}\BibitemShut {NoStop}%
\bibitem [{\citenamefont {Wadley}\ \emph {et~al.}(2016)\citenamefont {Wadley},
  \citenamefont {Howells}, \citenamefont {{\v{Z}}elezn{\`y}}, \citenamefont
  {Andrews}, \citenamefont {Hills}, \citenamefont {Campion}, \citenamefont
  {Nov{\'a}k}, \citenamefont {Olejn{\'\i}k}, \citenamefont {Maccherozzi},
  \citenamefont {Dhesi} \emph {et~al.}}]{wadley2016electrical}%
  \BibitemOpen
  \bibfield  {author} {\bibinfo {author} {\bibfnamefont {P.}~\bibnamefont
  {Wadley}}, \bibinfo {author} {\bibfnamefont {B.}~\bibnamefont {Howells}},
  \bibinfo {author} {\bibfnamefont {J.}~\bibnamefont {{\v{Z}}elezn{\`y}}},
  \bibinfo {author} {\bibfnamefont {C.}~\bibnamefont {Andrews}}, \bibinfo
  {author} {\bibfnamefont {V.}~\bibnamefont {Hills}}, \bibinfo {author}
  {\bibfnamefont {R.~P.}\ \bibnamefont {Campion}}, \bibinfo {author}
  {\bibfnamefont {V.}~\bibnamefont {Nov{\'a}k}}, \bibinfo {author}
  {\bibfnamefont {K.}~\bibnamefont {Olejn{\'\i}k}}, \bibinfo {author}
  {\bibfnamefont {F.}~\bibnamefont {Maccherozzi}}, \bibinfo {author}
  {\bibfnamefont {S.}~\bibnamefont {Dhesi}}, \emph {et~al.},\ }\bibfield
  {title} {\bibinfo {title} {Electrical switching of an antiferromagnet},\
  }\href@noop {} {\bibfield  {journal} {\bibinfo  {journal} {Science}\ }\textbf
  {\bibinfo {volume} {351}},\ \bibinfo {pages} {587} (\bibinfo {year}
  {2016})}\BibitemShut {NoStop}%
\bibitem [{\citenamefont {Bodnar}\ \emph {et~al.}(2018)\citenamefont {Bodnar},
  \citenamefont {{\v{S}}mejkal}, \citenamefont {Turek}, \citenamefont
  {Jungwirth}, \citenamefont {Gomonay}, \citenamefont {Sinova}, \citenamefont
  {Sapozhnik}, \citenamefont {Elmers}, \citenamefont {Kl{\"a}ui},\ and\
  \citenamefont {Jourdan}}]{bodnar2018writing}%
  \BibitemOpen
  \bibfield  {author} {\bibinfo {author} {\bibfnamefont {S.~Y.}\ \bibnamefont
  {Bodnar}}, \bibinfo {author} {\bibfnamefont {L.}~\bibnamefont
  {{\v{S}}mejkal}}, \bibinfo {author} {\bibfnamefont {I.}~\bibnamefont
  {Turek}}, \bibinfo {author} {\bibfnamefont {T.}~\bibnamefont {Jungwirth}},
  \bibinfo {author} {\bibfnamefont {O.}~\bibnamefont {Gomonay}}, \bibinfo
  {author} {\bibfnamefont {J.}~\bibnamefont {Sinova}}, \bibinfo {author}
  {\bibfnamefont {A.}~\bibnamefont {Sapozhnik}}, \bibinfo {author}
  {\bibfnamefont {H.-J.}\ \bibnamefont {Elmers}}, \bibinfo {author}
  {\bibfnamefont {M.}~\bibnamefont {Kl{\"a}ui}},\ and\ \bibinfo {author}
  {\bibfnamefont {M.}~\bibnamefont {Jourdan}},\ }\bibfield  {title} {\bibinfo
  {title} {Writing and reading antiferromagnetic mn2au by n{\'e}el spin-orbit
  torques and large anisotropic magnetoresistance},\ }\href@noop {} {\bibfield
  {journal} {\bibinfo  {journal} {Nature communications}\ }\textbf {\bibinfo
  {volume} {9}},\ \bibinfo {pages} {348} (\bibinfo {year} {2018})}\BibitemShut
  {NoStop}%
\bibitem [{\citenamefont {Shiino}\ \emph {et~al.}(2016)\citenamefont {Shiino},
  \citenamefont {Oh}, \citenamefont {Haney}, \citenamefont {Lee}, \citenamefont
  {Go}, \citenamefont {Park},\ and\ \citenamefont
  {Lee}}]{PhysRevLett.117.087203}%
  \BibitemOpen
  \bibfield  {author} {\bibinfo {author} {\bibfnamefont {T.}~\bibnamefont
  {Shiino}}, \bibinfo {author} {\bibfnamefont {S.-H.}\ \bibnamefont {Oh}},
  \bibinfo {author} {\bibfnamefont {P.~M.}\ \bibnamefont {Haney}}, \bibinfo
  {author} {\bibfnamefont {S.-W.}\ \bibnamefont {Lee}}, \bibinfo {author}
  {\bibfnamefont {G.}~\bibnamefont {Go}}, \bibinfo {author} {\bibfnamefont
  {B.-G.}\ \bibnamefont {Park}},\ and\ \bibinfo {author} {\bibfnamefont
  {K.-J.}\ \bibnamefont {Lee}},\ }\bibfield  {title} {\bibinfo {title}
  {Antiferromagnetic domain wall motion driven by spin-orbit torques},\ }\href
  {https://doi.org/10.1103/PhysRevLett.117.087203} {\bibfield  {journal}
  {\bibinfo  {journal} {Phys. Rev. Lett.}\ }\textbf {\bibinfo {volume} {117}},\
  \bibinfo {pages} {087203} (\bibinfo {year} {2016})}\BibitemShut {NoStop}%
\bibitem [{\citenamefont {Das}\ \emph {et~al.}(2023)\citenamefont {Das},
  \citenamefont {Cen}, \citenamefont {Wang},\ and\ \citenamefont
  {Fong}}]{PhysRevApplied.19.024063}%
  \BibitemOpen
  \bibfield  {author} {\bibinfo {author} {\bibfnamefont {D.}~\bibnamefont
  {Das}}, \bibinfo {author} {\bibfnamefont {Y.}~\bibnamefont {Cen}}, \bibinfo
  {author} {\bibfnamefont {J.}~\bibnamefont {Wang}},\ and\ \bibinfo {author}
  {\bibfnamefont {X.}~\bibnamefont {Fong}},\ }\bibfield  {title} {\bibinfo
  {title} {Bilayer-skyrmion-based design of neuron and synapse for spiking
  neural network},\ }\href {https://doi.org/10.1103/PhysRevApplied.19.024063}
  {\bibfield  {journal} {\bibinfo  {journal} {Phys. Rev. Appl.}\ }\textbf
  {\bibinfo {volume} {19}},\ \bibinfo {pages} {024063} (\bibinfo {year}
  {2023})}\BibitemShut {NoStop}%
\bibitem [{\citenamefont {Bradley}\ \emph {et~al.}(2024)\citenamefont
  {Bradley}, \citenamefont {Louis}, \citenamefont {Slavin},\ and\ \citenamefont
  {Tyberkevych}}]{bradley2024pattern}%
  \BibitemOpen
  \bibfield  {author} {\bibinfo {author} {\bibfnamefont {H.}~\bibnamefont
  {Bradley}}, \bibinfo {author} {\bibfnamefont {S.}~\bibnamefont {Louis}},
  \bibinfo {author} {\bibfnamefont {A.}~\bibnamefont {Slavin}},\ and\ \bibinfo
  {author} {\bibfnamefont {V.}~\bibnamefont {Tyberkevych}},\ }\bibfield
  {title} {\bibinfo {title} {Pattern recognition using spiking
  antiferromagnetic neurons},\ }\href@noop {} {\bibfield  {journal} {\bibinfo
  {journal} {Scientific Reports}\ }\textbf {\bibinfo {volume} {14}},\ \bibinfo
  {pages} {22373} (\bibinfo {year} {2024})}\BibitemShut {NoStop}%
\bibitem [{\citenamefont {Roldan}\ \emph {et~al.}(2022)\citenamefont {Roldan},
  \citenamefont {Maldonado}, \citenamefont {Aguilera-Pedregosa}, \citenamefont
  {Moreno}, \citenamefont {Aguirre}, \citenamefont {Romero-Zaliz},
  \citenamefont {Garc{\'\i}a-Vico}, \citenamefont {Shen},\ and\ \citenamefont
  {Lanza}}]{roldan2022spiking}%
  \BibitemOpen
  \bibfield  {author} {\bibinfo {author} {\bibfnamefont {J.~B.}\ \bibnamefont
  {Roldan}}, \bibinfo {author} {\bibfnamefont {D.}~\bibnamefont {Maldonado}},
  \bibinfo {author} {\bibfnamefont {C.}~\bibnamefont {Aguilera-Pedregosa}},
  \bibinfo {author} {\bibfnamefont {E.}~\bibnamefont {Moreno}}, \bibinfo
  {author} {\bibfnamefont {F.}~\bibnamefont {Aguirre}}, \bibinfo {author}
  {\bibfnamefont {R.}~\bibnamefont {Romero-Zaliz}}, \bibinfo {author}
  {\bibfnamefont {A.~M.}\ \bibnamefont {Garc{\'\i}a-Vico}}, \bibinfo {author}
  {\bibfnamefont {Y.}~\bibnamefont {Shen}},\ and\ \bibinfo {author}
  {\bibfnamefont {M.}~\bibnamefont {Lanza}},\ }\bibfield  {title} {\bibinfo
  {title} {Spiking neural networks based on two-dimensional materials},\
  }\href@noop {} {\bibfield  {journal} {\bibinfo  {journal} {npj 2D Materials
  and Applications}\ }\textbf {\bibinfo {volume} {6}},\ \bibinfo {pages} {63}
  (\bibinfo {year} {2022})}\BibitemShut {NoStop}%
\bibitem [{\citenamefont {Cao}\ \emph {et~al.}(2021)\citenamefont {Cao},
  \citenamefont {Meng}, \citenamefont {Chen}, \citenamefont {Liu},
  \citenamefont {Bian}, \citenamefont {Zhu}, \citenamefont {Liu},\ and\
  \citenamefont {Liu}}]{cao20212d}%
  \BibitemOpen
  \bibfield  {author} {\bibinfo {author} {\bibfnamefont {G.}~\bibnamefont
  {Cao}}, \bibinfo {author} {\bibfnamefont {P.}~\bibnamefont {Meng}}, \bibinfo
  {author} {\bibfnamefont {J.}~\bibnamefont {Chen}}, \bibinfo {author}
  {\bibfnamefont {H.}~\bibnamefont {Liu}}, \bibinfo {author} {\bibfnamefont
  {R.}~\bibnamefont {Bian}}, \bibinfo {author} {\bibfnamefont {C.}~\bibnamefont
  {Zhu}}, \bibinfo {author} {\bibfnamefont {F.}~\bibnamefont {Liu}},\ and\
  \bibinfo {author} {\bibfnamefont {Z.}~\bibnamefont {Liu}},\ }\bibfield
  {title} {\bibinfo {title} {2d material based synaptic devices for
  neuromorphic computing},\ }\href@noop {} {\bibfield  {journal} {\bibinfo
  {journal} {Advanced Functional Materials}\ }\textbf {\bibinfo {volume}
  {31}},\ \bibinfo {pages} {2005443} (\bibinfo {year} {2021})}\BibitemShut
  {NoStop}%
\bibitem [{\citenamefont {Ovcharov}\ \emph {et~al.}(2022)\citenamefont
  {Ovcharov}, \citenamefont {Galkina}, \citenamefont {Ivanov},\ and\
  \citenamefont {Khymyn}}]{PhysRevApplied.18.024047}%
  \BibitemOpen
  \bibfield  {author} {\bibinfo {author} {\bibfnamefont {R.}~\bibnamefont
  {Ovcharov}}, \bibinfo {author} {\bibfnamefont {E.}~\bibnamefont {Galkina}},
  \bibinfo {author} {\bibfnamefont {B.}~\bibnamefont {Ivanov}},\ and\ \bibinfo
  {author} {\bibfnamefont {R.}~\bibnamefont {Khymyn}},\ }\bibfield  {title}
  {\bibinfo {title} {Spin hall nano-oscillator based on an antiferromagnetic
  domain wall},\ }\href {https://doi.org/10.1103/PhysRevApplied.18.024047}
  {\bibfield  {journal} {\bibinfo  {journal} {Phys. Rev. Appl.}\ }\textbf
  {\bibinfo {volume} {18}},\ \bibinfo {pages} {024047} (\bibinfo {year}
  {2022})}\BibitemShut {NoStop}%
\bibitem [{\citenamefont {Chotorlishvili}\ and\ \citenamefont
  {Ugulava}(2010)}]{chotorlishvili2010quantum}%
  \BibitemOpen
  \bibfield  {author} {\bibinfo {author} {\bibfnamefont {L.}~\bibnamefont
  {Chotorlishvili}}\ and\ \bibinfo {author} {\bibfnamefont {A.}~\bibnamefont
  {Ugulava}},\ }\bibfield  {title} {\bibinfo {title} {Quantum chaos and its
  kinetic stage of evolution},\ }\href@noop {} {\bibfield  {journal} {\bibinfo
  {journal} {Physica D: Nonlinear Phenomena}\ }\textbf {\bibinfo {volume}
  {239}},\ \bibinfo {pages} {103} (\bibinfo {year} {2010})}\BibitemShut
  {NoStop}%
\bibitem [{\citenamefont {Ugulava}\ \emph {et~al.}(2003)\citenamefont
  {Ugulava}, \citenamefont {Chotorlishvili},\ and\ \citenamefont
  {Nickoladze}}]{PhysRevE.68.026216}%
  \BibitemOpen
  \bibfield  {author} {\bibinfo {author} {\bibfnamefont {A.}~\bibnamefont
  {Ugulava}}, \bibinfo {author} {\bibfnamefont {L.}~\bibnamefont
  {Chotorlishvili}},\ and\ \bibinfo {author} {\bibfnamefont {K.}~\bibnamefont
  {Nickoladze}},\ }\bibfield  {title} {\bibinfo {title} {Overlapping of
  nonlinear resonances and the problem of quantum chaos},\ }\href
  {https://doi.org/10.1103/PhysRevE.68.026216} {\bibfield  {journal} {\bibinfo
  {journal} {Phys. Rev. E}\ }\textbf {\bibinfo {volume} {68}},\ \bibinfo
  {pages} {026216} (\bibinfo {year} {2003})}\BibitemShut {NoStop}%
\bibitem [{\citenamefont {Ugulava}\ \emph {et~al.}(2004)\citenamefont
  {Ugulava}, \citenamefont {Chotorlishvili},\ and\ \citenamefont
  {Nickoladze}}]{PhysRevE.70.026219}%
  \BibitemOpen
  \bibfield  {author} {\bibinfo {author} {\bibfnamefont {A.}~\bibnamefont
  {Ugulava}}, \bibinfo {author} {\bibfnamefont {L.}~\bibnamefont
  {Chotorlishvili}},\ and\ \bibinfo {author} {\bibfnamefont {K.}~\bibnamefont
  {Nickoladze}},\ }\bibfield  {title} {\bibinfo {title} {Quantum-mechanical
  research on nonlinear resonance and the problem of quantum chaos},\ }\href
  {https://doi.org/10.1103/PhysRevE.70.026219} {\bibfield  {journal} {\bibinfo
  {journal} {Phys. Rev. E}\ }\textbf {\bibinfo {volume} {70}},\ \bibinfo
  {pages} {026219} (\bibinfo {year} {2004})}\BibitemShut {NoStop}%
\bibitem [{\citenamefont {Ugulava}\ \emph {et~al.}(2005)\citenamefont
  {Ugulava}, \citenamefont {Chotorlishvili},\ and\ \citenamefont
  {Nickoladze}}]{PhysRevE.71.056211}%
  \BibitemOpen
  \bibfield  {author} {\bibinfo {author} {\bibfnamefont {A.}~\bibnamefont
  {Ugulava}}, \bibinfo {author} {\bibfnamefont {L.}~\bibnamefont
  {Chotorlishvili}},\ and\ \bibinfo {author} {\bibfnamefont {K.}~\bibnamefont
  {Nickoladze}},\ }\bibfield  {title} {\bibinfo {title} {Irreversible evolution
  of quantum chaos},\ }\href {https://doi.org/10.1103/PhysRevE.71.056211}
  {\bibfield  {journal} {\bibinfo  {journal} {Phys. Rev. E}\ }\textbf {\bibinfo
  {volume} {71}},\ \bibinfo {pages} {056211} (\bibinfo {year}
  {2005})}\BibitemShut {NoStop}%
\end{thebibliography}%

\end{document}